\documentclass[11pt]{article}
		\usepackage{verbatim,amsmath,amssymb}
		\usepackage{epsfig,float,color}
		\usepackage{epstopdf}
		\usepackage{geometry}
		\usepackage{setspace}
		\usepackage{wrapfig}
		\usepackage{hyperref}
		\usepackage[utf8]{inputenc}
		\usepackage{graphicx}
		\usepackage{flushend}
		\usepackage[linesnumbered,ruled]{algorithm2e}
	    
		\usepackage[font={footnotesize}]{caption}
		\usepackage[font={footnotesize}]{subcaption}
		\usepackage{footnote}
		\usepackage{multicol}
		\usepackage{multirow}
		\usepackage{enumitem}   
		\usepackage{soul}
		\usepackage{framed}
		\usepackage[titletoc,title]{appendix}
		\usepackage{bm}
		\usepackage{cite}
		\usepackage[makeroom]{cancel} 
		\definecolor{darkblue}{rgb}{0,0,1}
		\hypersetup{pdftex=true, colorlinks=true, breaklinks=true, linkcolor=darkblue, menucolor=darkblue, pagecolor=darkblue, citecolor=darkblue, urlcolor=darkblue}
		
		\usepackage[dvipsnames]{xcolor}
\begin{document}
		
		\begin{center}
		    \Large{\bf{Compliant Fluidic Control Structures: Concept and Synthesis Approach}}\\
			
		\end{center}
		
			\begin{center}
			\large{Prabhat Kumar\footnote{corresponding author: p.kumar-3@tudelft.nl,\,prabhatkumar.rns@gmail.com}, Paola Fanzio, Luigi Sasso, and Matthijs Langelaar\footnote{Email: M.Langelaar@tudelft.nl}}
			\vspace{4mm}
			
			\small{\textit{Department of Precision and Microsystems Engineering, Faculty of 3mE, Delft University of
					Technology, Mekelweg 2, 2628 CD Delft, the Netherlands}}
			
			\vspace{4mm}
			
			Published\footnote{This pdf is the personal version of an article whose final publication is available at \href{https://www.sciencedirect.com/science/article/pii/S0045794918315232}{www.sciencedirect.com}} 
			in \textit{Computers \& Structures}, 
			\href{https://doi.org/10.1016/j.compstruc.2019.02.004}{DOI:10.1016/j.compstruc.2019.02.004} \\
			Submitted on 29.~October 2018, Revised on 22. January 2019, Accepted on 20. February 2019
			
		\end{center}
		
		\vspace{3mm}
		\rule{\linewidth}{.15mm}
		{\bf Abstract:}
			The concept and synthesis approach for planar Compliant Fluidic Control Structures (CFCSs), monolithic flexible continua with embedded functional pores, is presented in this manuscript. Such structures are envisioned to find application in biomedicine as tunable microfluidic devices for drug/nutrient delivery. The functional pores enlarge and/or contract upon deformation of the compliant structure in response to external stimuli, facilitating the regulated control of fluid/nutrient/drug transport. A thickness design variable based topology optimization problem is formulated to generate effective designs of these structures. An objective based on hydraulic diameter(s) is conceptualized, and it is extremized using a gradient based optimizer. 
			 Both geometrical and material nonlinearities are considered. The nonlinear behaviour of employed hyperelastic  material is modeled via the Arruda-Boyce constitutive material model. Large-displacement finite element analysis is performed using the updated Lagrangian formulation in plane-stress setting. The proposed synthesis approach is applied to various CFCSs for a variety of fluidic control functionalities. The optimized designs of various CFCSs with single and/or multiple functional pores are fabricated via a Polydimethylsiloxane (PDMS) soft lithography process, using a high precision 3D printed mold and their performances are compared with the numerical predictions. \\
			 

		\vspace{-4mm}
		\rule{\linewidth}{.15mm}
		

	\section{INTRODUCTION}\label{I}
		
		A compliant mechanism (CM) is a monolithic structure which performs a task by deriving a part or whole of its relative motion from small/large elastic deformation of its constituting flexible members. By virtue of flexible monolithic design/flexible parts, such mechanisms possess many advantages over conventional linkage-type mechanisms, e.g., low manufacturing/assembly cost, less frictional losses and less wear/tear due to the absence of kinematic joints, low maintenance cost, high precision, repeatability, and scalability, to name a few \cite{howell2001compliant}. Therefore, the use of CMs is on rise in various fields which span from  simple house clothespins to applications requiring highly precise motion/deformation characteristics, e.g., medical instruments \cite{dziedzic2001design}, MEMS \cite{ananthasuresh1994strategies}, adaptive structures \cite{saggere1999static}, path generation \cite{saxena2001topology,pedersen2001topology}, \textit{etc}. In this manuscript, we present the concept and  synthesis approach for \textit{Compliant Fluidic Control Structures (CFCSs)}. Such structures, monolithic flexible continua with embedded functional pores, can find application in biomedicine as tunable devices for delivering nutrients/drugs to cells. The embedded functional pores are the regions which facilitate regulation of fluid/nutrient/drug transport by enlarging and/or contracting in response to external loading(s). For the intended applications, one can ultimately target typical controllable pore sizes between 5 to 50 micrometers.  Herein, we primarily focus on the synthesis approach, and evaluate the results on larger-scale demonstrator devices for convenience.
		
		Microfluidic cell-, tissue- and organ-on-chip systems have shown the potential to revolutionize biomedicine and drug discovery wherein tunable devices are required \cite{fanzio2012modulating,esch2015organs}. 
		However, designing such devices which can render high-precision control over localized transport of nutrient and drugs, is challenging. In addition, from a manufacturing point of view, it is difficult to integrate local actuators near the functional pores within the device to achieve the desired fluid flow in response to control inputs. Therefore, instead of integrating local actuators, our focus is on achieving desired localized effects by remote actuation through a suitable mechanical transfer mechanism. It may be possible to design such tunable microfluidic devices intuitively, however, obtaining high-performance designs with multiple functional pores, may require numerous iterations of trial and error, and does not offer systematic solution procedures. As an alternative, we use topology optimization to generate such devices by extremizing the formulated objective with the given resource constraints (if any). The embedded functional pores of a CFCS either are enlarged or contracted in response to the input loadings wherein the structure can experience large deformation. To give an indication, the devices designed as examples in this paper are subject to strains up to 100\%.
		
		  Polydimethylsiloxane (PDMS) elastomer material is widely used to fabricate microfluidic devices since it offers various advantages \cite{mcdonald2000fabrication} including adequate optical transparency, low cost for mass production using, e.g., lithographic techniques, bio-compatibility, and permitting large deformation. Because of these properties, we use PDMS material to fabricate the optimized CFCS continua. Mechanically, PDMS exhibits characteristics similar to rubber-like materials \cite{lotters1997mechanical}. Therefore, it is essential to consider material nonlinearity in the synthesis approach. Herein, the material definition proposed by Arruda and Boyce \cite{arruda1993three,holzapfel2001nonlinear} is employed to model the behavior of the PDMS. 
		 
		Topology optimization (TO) determines an optimized material layout within a given design domain $\mathcal{B}_0$ with known boundary conditions by extremizing the formulated objective(s) for the desired output response(s) under the specified constraint(s) (if any). Each TO iteration solves the associated boundary value problem(s). In the considered structural optimization context, typically, a ubiquitous option is to use finite element (FE) analysis wherein one can either employ discrete (beam/frame) or continuous (quad/triangular/polygonal) FEs to parameterize $\mathcal{B}_0$. There exist many approaches \cite{martin2003topology} which consider linear FE analysis to achieve the optimized design for a wide range of single/multi-physics mechanical problems. Numerous approaches involving TO have been presented to synthesize CMs. Normally, these methods find a trade-off between stiffness measures (e.g., compliance/strain energy) and flexibility measures (e.g., output displacements/mutual strain energy) of the compliant continua \cite{deepak2009comparative}. On the other hand, TO approaches involving large deformation with/without nonlinear constitutive models \cite{buhl2000stiffness,bruns2001topology,yoon2005element,van2014element,langelaar2011topology,saxena2013combined,Klarbring2013,Lahuerta2013,wang2014interpolation,luo2015topology,liu2017design} are much less common. This could be mostly due to numerical difficulties in handling significantly large deformation characterized via geometrical and material nonlinearities and distortion/inversion of low-stiffness FEs. 
		
		To avoid the numerical instabilities in large deformation TO, the approach in \cite{buhl2000stiffness} did not regard the internal forces originating at nodes surrounded by low-stiffness elements in its Newton-Raphson convergence criteria.  The method in \cite{bruns2001topology} treated such instabilities by removing and reintroducing low-stiffness elements during optimization. Yoon and Kim \cite{yoon2005element} proposed connectivity parameterization using fictitious springs and Langelaar et al. \cite{langelaar2011topology} used this approach to design planar shape memory alloy thermal actuators experiencing large deformation. Van Dijk et al. \cite{van2014element} presented an element deformation scaling approach. Refs. \cite{buhl2000stiffness,bruns2001topology,yoon2005element,van2014element} considered a St. Venant-Kirchhoff material model in their approaches. Saxena and Sauer \cite{saxena2013combined} combined zeroth and first order optimization techniques wherein the material nonlinearity was modeled using a neo-Hookean constitutive model. Wang et al. \cite{wang2014interpolation} proposed an interpolation scheme to deal with distortion/inversion of low-stiffness elements and they solved problems using both  St. Venant-Kirchhoff and neo-Hookean material models. Noting the fact that a St. Venant-Kirchhoff material model fails to provide the actual response in a large compression regime, Lahuerta et al. \cite{Lahuerta2013} and Klarbring et al. \cite{Klarbring2013} employed  relatively more realistic hyperelastic material models in their large-deformation TO approaches.  Luo et al. \cite{luo2015topology} proposed a method using an additive hyperelasticity technique. To generate CMs experiencing large deformation, Liu et al. \cite{liu2017design} presented a modified additive hyperelaticity based approach. Presumably, the large-deformation CFCS optimization is more robust than general large-deformation TO because the stiffness differences that occur in the structure are smaller in our case due to manufacturing restrictions (geometrical construction) and thus, no special techniques are needed to reach a stable optimization result. This is discussed in more detail in Section \ref{section:2}.
		
	   This paper presents an approach using TO to synthesize planar Compliant Fluidic Control Structures. To provide a proper treatment of large deformation of the CFCSs (fabricated using PDMS), both geometrical and material nonlinearities are incorporated within the approach. In addition, as planar structures are targeted in this paper, \textit{plane-stress} conditions are explicitly imposed within the nonlinear FE formulation using the method described in  \cite{klinkel2002using,zienkiewicz2005finite}.  An objective based on hydraulic diameter(s) of the pore(s) is formulated and extremized to achieve the desired performances, i.e., enlarging and/or contracting of the functional pores to facilitate the required fluid flow. Depending upon the various applications, there can be single/multiple embedded functional pores within a CFCS.  CFCSs with multiple pores can facilitate diverse tailored fluid flow characteristics in response to a single remote actuation. Note that the effect of fluid flow is not incorporated while optimizing the CFCSs since in the intended applications flow rates and pressure differences are small. The design of other compliant structures where fluid-structure interaction becomes relevant forms an additional challenge beyond the scope of this work. In this paper, we focus on planar CFCSs, in view of the application demands and the available manufacturing options. However, in principle the concept can be readily extended to a general 3D TO setting.
		
		In summary, the new contributions of the current work are:
		\begin{itemize}
			\item the concept and synthesis approach for Compliant Fluidic Control Structures which can find application in, e.g., biomedicine/drug delivery as tunable devices to facilitate precise transport of nutrients and/or drugs,
			\item formulation of a TO-based synthesis approach using thickness design variables \cite{ROSSOW1973} while considering geometrical and material nonlinearities under  \textit{plane-stress} conditions,
			\item an objective based on hydraulic diameter is conceptualized and minimized to achieved the desired modes (enlargement/contraction) of the functional pores,
			\item use of the Arruda-Boyce material definition \cite{arruda1993three,holzapfel2001nonlinear} to model nonlinear mechanical behaviour of PDMS in the TO setting,
			\item demonstration of the approach by synthesizing various CFCSs having single and/or multiple functional pores for achieving a variety of fluid flow control scenario,  
			\item a flexible, low cost fabrication technique using PDMS material for the optimized CFCS continua,
			\item comparison of the performances of the fabricated CFCFs with their respective numerical models.
		\end{itemize}

		The remainder of the paper is organized as follows. Section \ref{section:2} presents the concept and methodology involving problem definition and formulation, fabrication technique and experimental setup, PDMS material modeling and the numerical technique. The objective formulation and sensitivity analysis is given in Section \ref{section:3}. Section \ref{section:4} presents the numerical results and performance comparison for numerical models and respective experimental counterparts. Lastly, conclusions are drawn in Section \ref{section:5}.     
	
		\section{CONCEPT and METHODOLOGY}\label{section:2}
		This section describes problem definition and formulation, fabrication technique and experimental setup, material modeling for PDMS and the employed numerical technique.
		
\subsection{Problem definition and formulation}\label{PDAF}
\begin{figure}[h!]
					\begin{subfigure}[t]{0.450\textwidth}
						\centering
						\includegraphics[scale=1]{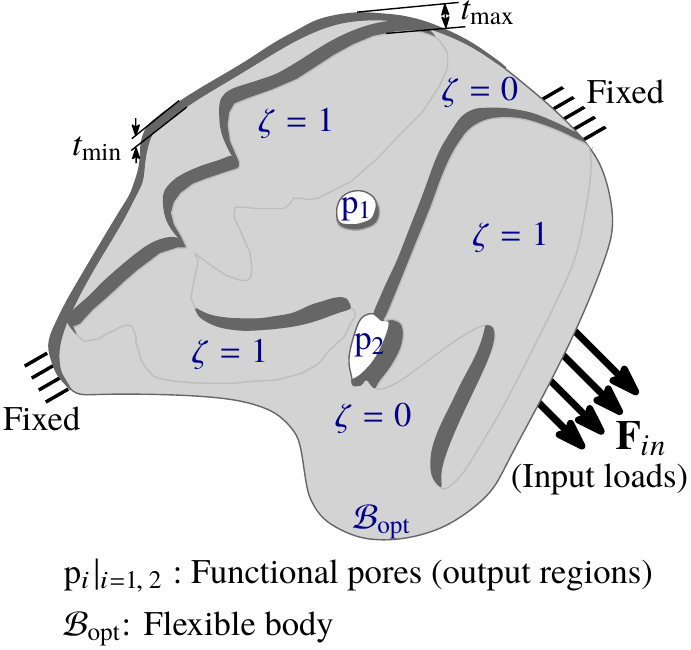}
						\caption{Planar compliant fluidic control structure}
						\label{fig:fig1a}
					\end{subfigure}
					~ \qquad
					\begin{subfigure}[t]{0.45\textwidth}
						\centering
						\includegraphics[scale=1]{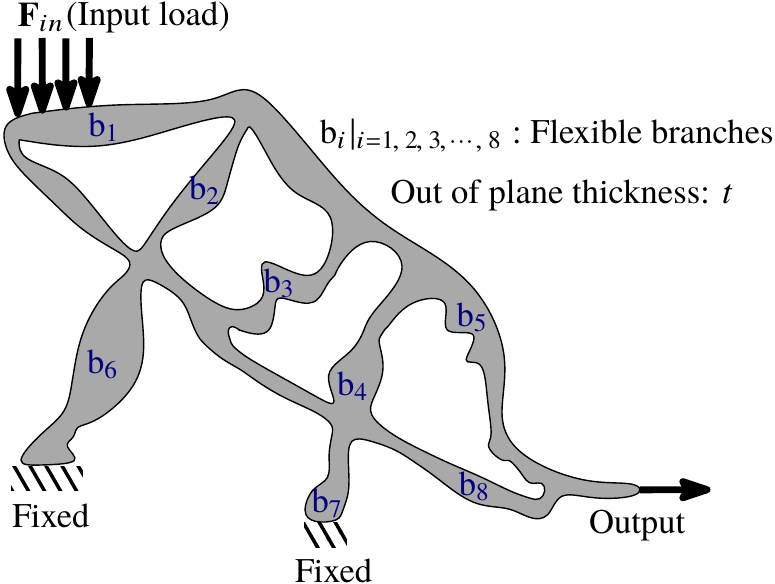}
						\caption{Planar compliant mechanism}
						\label{fig:fig1b}
					\end{subfigure}
					\caption{Geometrical construction of a CFCS and a compliant mechanism with fixed boundary conditions, input loads $\mathbf{F}_{in}$ and respective output locations are depicted. A planar CFCS (Fig. \ref{fig:fig1a}) has varying thickness (minimum $t_{\min}$ and maximum $t_{\max}$) and it is filled with the actual material in its entire design domain $\mathcal{B}_\mathrm{out}$. The parameter $\zeta$ defines the thickness at the different locations for the CFCS. $\zeta =0$ indicates the region having  thickness $t_{\min}$, and the regions with $\zeta =1$ possess thickness  $t_{\max}$. Two embedded functional pores p$_1$ and p$_2$ can be enlarged and/or contracted in response to external force $\mathbf{F}_{in}$ and thus, regulate the desired transport of nutrient and/or drugs. The entire body $\mathcal{B}_\mathrm{out}$ of the CFCS is flexible. In case of compliant mechanisms (Fig. \ref{fig:fig1b}), material is relocated in specific manner with uniform thickness $t$ throughout the optimized layout (in gray) which is constituted of flexible branches b$_i|_{i= 1,\,2,\,3,\cdots,\,8}$.}\label{fig:fig1}
				\end{figure}

Regarding the geometrical features, a planar CFCS is different from a planar CM. The former has actual material throughout its (optimized) design domain $\mathcal{B}_\mathrm{opt}$ with different topographical features (different thickness at different locations) with embedded functional pores as shown in Fig. \ref{fig:fig1a}, however, the latter has uniform thickness wherever the actual material is designated into the connected finite regions (branches) via the synthesis approach (Fig. \ref{fig:fig1b}). In other words, in a planar setting, a CFCS is a monolithic flexible continuum with functional pore(s) having different topographical features and a CM, in general, is constituted via a particular topology consisting of slender flexible branches. Note that the shape and distribution of the thicker regions in a planar CFCS can also be seen to constitute a topology. Both perform their tasks using the motion obtained from the (large) deformation in their respective structures. The regions associated to different topographical features of a CFCS will undergo different deformation characteristics in response to external stimuli and thus, help achieving the desired tasks of the functional pores. Evidently, for a CFCS, specific deformation behaviour of the boundary defining the pores is desired, however in case of a CM, generally, interest lies in the deformation characteristics of a single output point.

    	To generate the different topographical features available in the geometrical construction of CFCSs via TO, an alternative design parameterization involving thickness is proposed, which is termed \textit{thickness design variable} $\zeta$, herein. In the discrete (FE) setting, a thickness design variable  $\zeta_e \in [0,\,\, 1]$ is designated to each element.  These variables do not govern the material states \cite{martin2003topology} of the element, but instead determine its out-of-plane thickness during optimization. The lower and upper limits of achievable (desired) thickness are represented via $t_\mathrm{\min}$ and $t_\mathrm{\max}$, respectively. Then for a CFCS design, finite regions/elements with $\zeta_e = 0$ will have thickness $t_\mathrm{\min}$ and those with $\zeta_e= 1$ indicate thickness $t_\mathrm{\max}$ (Fig. \ref{fig:fig1}). Herein, the ratio $\frac{t_\mathrm{\min}}{t_\mathrm{\max}}$ is chosen in perspective of the applied fabrication technique which limits the $t_{\min}$ to 0.5 mm. Note, using an FE setting for a planar geometry(ies), i.e., in a plane-stress case, one can define internal forces and hence, material and geometrical stiffness matrices, as a function of thickness \cite{ROSSOW1973,zienkiewicz2005finite} (see Appendix \ref{Append}). Indeed, this proposed discretization helps to formulate the problem as a variable sheet thickness problem \cite{ROSSOW1973}, and lead us to define thickness design variable $\zeta_e$ for each element as a natural design variable. In addition, from the application point of view, new holes should not be introduced during optimization. Therefore, the approach can been seen as \textit{topography optimization} instead of topology optimization.

		  When using gradient based optimization, sensitivities of the objective and constraints with respect to the design variables are required.  In that regard, the thickness of each element is varied continuously as a function of the design variable $\zeta_e$ as: 
		   \begin{equation}\label{Eq:1}
		   t_e(\zeta_e) = t_\mathrm{\min} + \zeta_e^p(t_\mathrm{\max} - t_\mathrm{\min}), \qquad \zeta_e\in[0,\,1]
		   \end{equation}
		 where $p$ is the penalization factor which steers the optimization towards either $t_\mathrm{\min}$ or $t_\mathrm{\max}$. One  notices, the formulation (Eq. \ref{Eq:1}) is similar to the classical SIMP formulation \cite{martin2003topology,rozvany1992generalized} of TO.  It is not required to relax the lower limit of the design variables during the optimization procedure since when $\zeta_e$ equals to zero then $t_e$ and thus, elemental stiffness matrices (Eq. \ref{Eq:1} and Appendix \ref{Append}) remain nonzero \cite{zienkiewicz2005finite}. Therefore, numerical instability does not arise, if $\zeta_e$ is not relaxed while ${t_\mathrm{\min}}$ is set to a finite value in view of the fabrication technique. In addition, throughout the optimization process elements with $\zeta_e =0$ maintain proper stiffnesses and thus, demand no special treatments for large deformation TO \cite{wang2014interpolation}. 
		  
		The general CFCS optimization problem is formulated as
		\begin{equation}\label{Eq:2}
			\begin{aligned}
				& \underset{\bm{\zeta}}{\text{min}}
				& &f_0(\bm{\zeta}) \\
				& \text{such that} & & \mathbf{R}(\mathbf{u},\bm{\zeta}) = \mathbf{0} \\
				&  && V(\bm{\zeta})-V^*\le 0
			\end{aligned}
		\end{equation}
		where $f_0(\bm{\zeta})$ (Section \ref{section:3}) is the objective to be minimized, $\bm{\zeta}$ is the design vector consisting of all design variables $\zeta_e|_{e = 1,\,2,\,3,\,\cdots ,\,N_\mathrm{elem}}$, $\mathbf{R}(\mathbf{u},\bm{\zeta})$ is the residual force vector stemming from the mechanical equilibrium equations introduced below (Eq. \ref{Eq:17}), and the current and permitted volume are represented via  $V(\bm{\zeta})\,\text{and}\,V^*$, respectively. Further, $\mathbf{u}$ is the displacement vector and $N_\mathrm{elem}$ is the total number of FEs used to parametrize the design domain $\mathcal{B}_0$. Herein, a constraint on resource volume is imposed so that the required functionalities of CFCSs can be achieved using the material within the given limits. However, if not desired, volume constraints can be relaxed and hence, the optimizer can select the required material to be used automatically. 
		
		We implement the density filter \cite{bruns2001topology,bourdin2001filters} to ensure a minimum length scale of the structural features for manufacturing convenience. The release step of the PDMS molding process used for the fabricated samples, as described below (Section \ref{FT}), does not allow very small features. In view of the filter, the physical variable $\tilde{\zeta_e}$ is evaluated from the original design variable $\zeta_e$ as
		\begin{equation}\label{Eq:3}
		\tilde{\zeta_e} = \frac{1}{\displaystyle\sum_{k\in N_e}w_{ek}}\displaystyle\sum_{k \in N_e} w_{ek}\zeta_k
		\end{equation}
		where $N_e = \{k,\,||\mathbf{x}_k^c - \mathbf{x}_e^c||\le r_{\mathrm{\min}}\}$. Herein, $\mathbf{x}_k^c\,\text{and}\, \mathbf{x}_e^c$ denote the center coordinates of the $k^\mathrm{th}$ and $e^\mathrm{th}$ elements respectively, and $r_\mathrm{\min}$ is the user defined filter radius. Further, $w_{ek}$ represents weight function which is defined as
		\begin{equation}\label{Eq:4}
		w_{ek} = \max\{0, r_\mathrm{\min} - ||\mathbf{x}_k^c - \mathbf{x}_e^c||\}.
		\end{equation}
		We use MMA \cite{svanberg1987method}, a gradient-based optimizer, to perform the TO. 
\subsection{Fabrication Technique and Experimental setup}	\label{FT}
\begin{figure}[h!]
	\begin{subfigure}[t]{0.250\textwidth}
		\centering
		\includegraphics[scale=0.5]{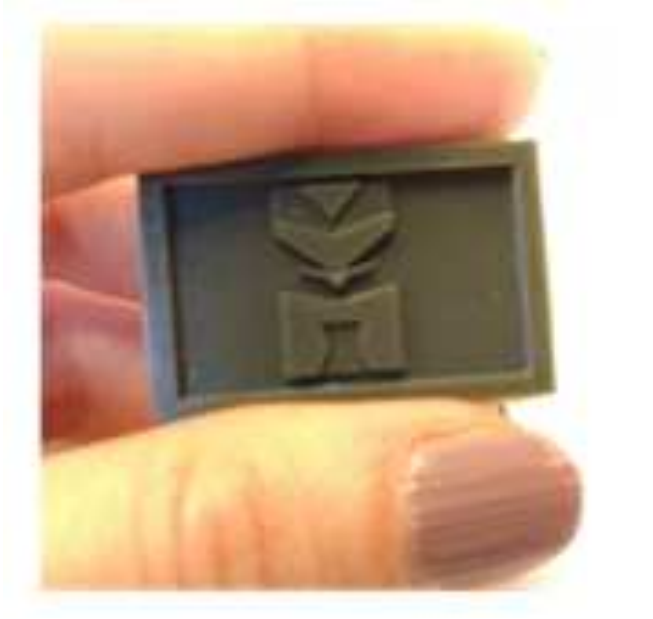}
		\caption{}
		\label{fig:fig2a}
	\end{subfigure}
	~ 
	\begin{subfigure}[t]{0.25\textwidth}
		\centering
		\includegraphics[scale=0.5]{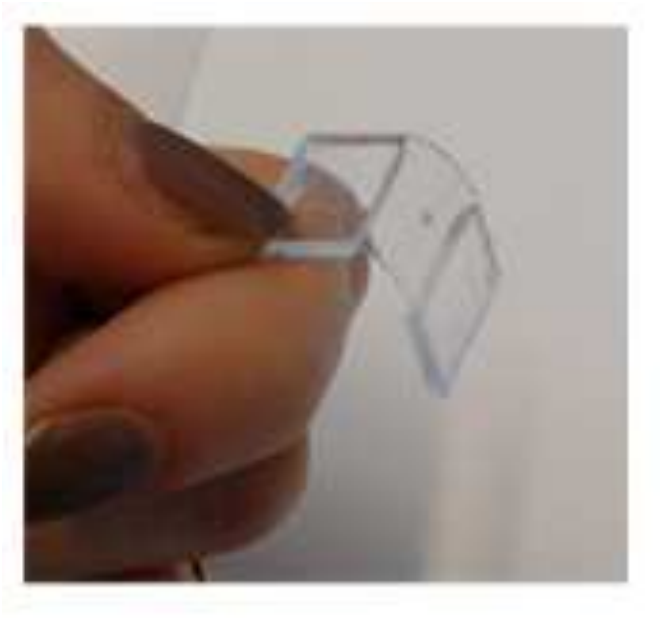}
		\caption{}
		\label{fig:fig2b}
	\end{subfigure}
	\begin{subfigure}[t]{0.5\textwidth}
		\centering
		\includegraphics[scale=0.5]{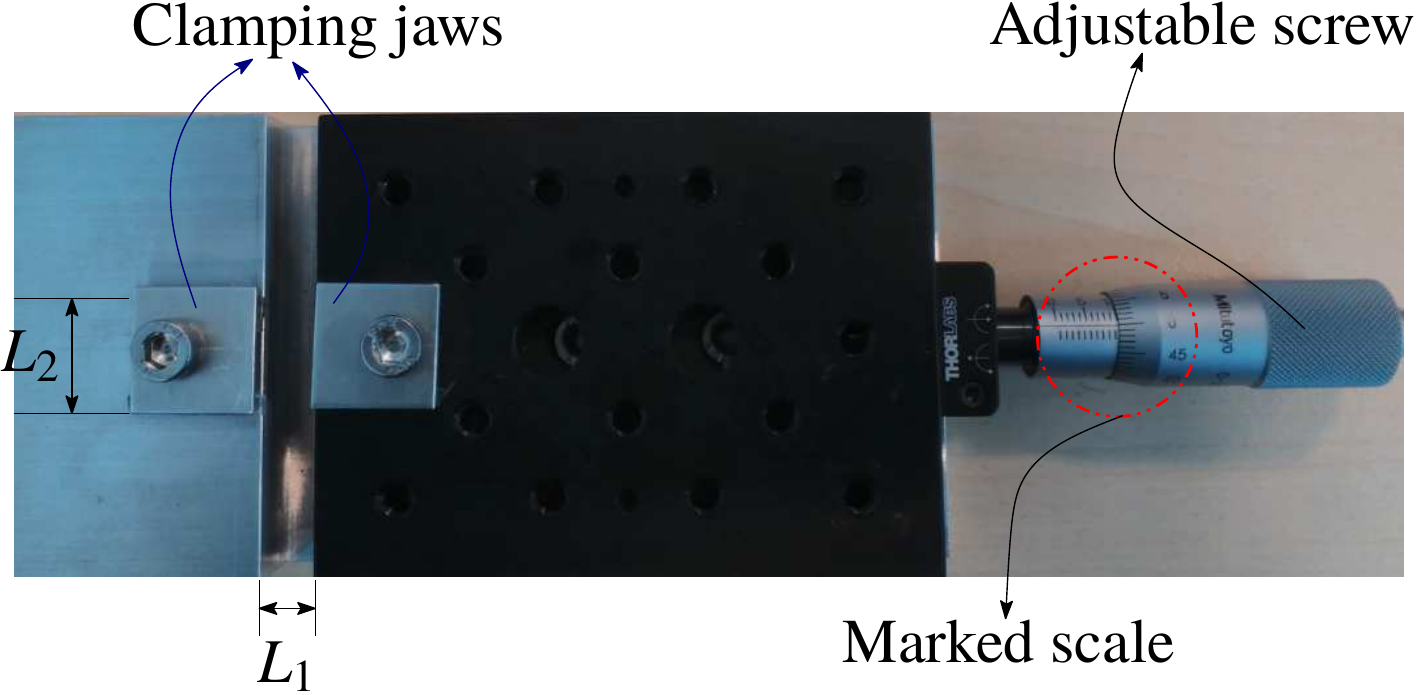}
		\caption{}
		\label{fig:fig2c}
	\end{subfigure}
	\caption{(\subref{fig:fig2a}) 3D-printed mold, (\subref{fig:fig2b}) A Compliant Fluidic Control Structure (CFCS) fabricated by a PDMS molding process, and (\subref{fig:fig2c}) An equipment used to perform the experiment, \textbf{Key}: $L_1 = 10$ mm and $L_2 = 16$ mm} \label{fig:fig2pre}
\end{figure}

	A replica molding process \cite{fanzio2011dna} is used to fabricate the optimized CFCSs wherein first, the structures are translated into respective mold designs. PDMS is then cast in the mold to obtain the final prototypes. The following steps are adopted in the fabrication process:
\begin{enumerate}
	\item The optimized design is translated into a CAD model to design a corresponding mold. Defeaturing is applied where necessary to prevent small structures to remain stuck in the mold during release.
	\item The mold is fabricated via a stereolithographic 3D printing process (Envisiontec Micro Plus Hires), layer thickness $25\,\mu$m, by using the HTM $140$ polymer (Fig. \ref{fig:fig2a}).
	\item The mold is then cleaned for two minutes using IPA in an ultrasound bath, dried and placed under UV light (Photopol light, Dentalfarm) for another two minutes.
	\item PDMS (1:10) is then mixed and degassed for thirty minutes to remove air bubbles. 
	\item PDMS is then casted into the mold and the excess is removed with a flat glass coverslip.
	\item Thereafter, PDMS is cured for one hour at 70$^o$C and subsequently removed from the mold obtaining the final prototype of the CFCS (Fig. \ref{fig:fig2b}). 
\end{enumerate}
		
			Figure \ref{fig:fig2c} depicts the experimental setup. Two clamping jaws can be tightened to keep the membrane in place. Slipping of the PDMS membrane (CFCS) is avoided using two protrusions that increase the friction. An adjustable screw facilitates for applying a desired stretching deformation to the membrane using the marked scale. Note that in view of the experimental setup and manufacturing limitations, we consider the design domain of size $L_1\times L_2\,(= 10\times 16)$ mm$^2$ with embedded pores of size $0.5\times 0.5$ mm$^2$ to generate CFCSs using TO.
		
		\subsection{Material Modeling for PDMS}\label{PDMSMM}
		In this section, we briefly review kinematics of deformation for the sake of completeness and to introduce necessary terminology. Thereafter, the material model proposed by Arruda and Boyce \cite{arruda1993three,holzapfel2001nonlinear} which is used herein to model the behavior of PDMS, is presented. 
			 
		\subsubsection{Kinematics of Deformation}
\begin{wrapfigure}{r}{0.5\textwidth}
		 	\centering
		 		\includegraphics[width=0.48\textwidth]{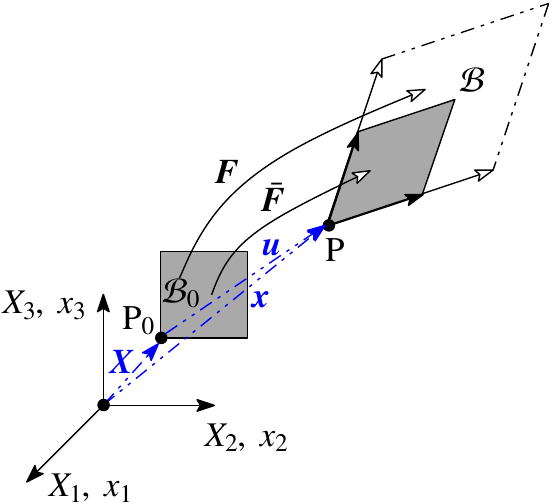}
		 	\caption{A schematic diagram for dilatation and distortional deformation. $\bar{\bm{F}}$ is the deformation gradient for isochoric deformation, i.e., $\det \bar{\bm{F}}= 1$ and hence, $\bar{\bm{F}} = J^{-1/3}\bm{F}$.} \label{fig:fig3}
		 \end{wrapfigure}

		Typically, the deformation gradient $\bm{F}$ is used to described finite deformation of a body $\mathcal{B}_0$ (Fig. \ref{fig:fig3}). It relates physical quantities before deformation to the corresponding quantities after/during deformation, i.e., 
		\begin{equation}\label{Eq:5}
		\bm{F}=\frac{\partial \bm{x}}{\partial \bm{X}} = \nabla_0\bm{u} + \bm{I} = \begin{bmatrix}
		F_{11} & F_{12} &  F_{13}\\
		F_{21} & F_{22} &  F_{23}\\
		F_{31} & F_{32} &  F_{33}\\
		\end{bmatrix} \text{(in matrix notation)}
		\end{equation}
		where $\bm{X}\in\mathcal{B}_0$ is the reference configuration state of a material point P$_0$, say. $\bm{x}\in\mathcal{B}$ and $\bm{u} = \bm{x}-\bm{X}$ are the corresponding spatial configuration state and the displacement, respectively. Further, $\bm{I}$ is the identity tensor and  $\nabla_0\bm{u}$ represents the gradient of $\bm{u}$ with respect to the reference configuration $\bm{X}$.
		 
		In terms of the deformation gradient $\bm{F}$, one defines the left and right Cauchy-Green deformation tensors $\bm{B}$ and $\bm{C}$ as
		\begin{equation}\label{Eq:6}
		\bm{B}=\bm{F}\bm{F}^T,\qquad\bm{C}= \bm{F}^T\bm{F}.
		\end{equation} 
		As per the polar decomposition theorem \cite{holzapfel2001nonlinear}, $\bm{F}$ can be written using the right stretch tensor $\bm{U}$ or the left stretch tensor $\bm{V} (= \bm{RU}\bm{R}^T)$ and a rotation tensor $\bm{R}$ as, $\bm{F} = \bm{R}\bm{U}= \bm{VR}$. Let the eigenvalues (principal stretches) and eigenvectors (principal directions)  of the right stretch tensor $\bm{U}$ be  $\lambda_i|_{i=1,\,2,\,3}$ and $\bm{Y}_i$, respectively. Then using the spectral decomposition theorem and noting the fact that the right Cauchy-Green deformation tensors $\bm{C}$ have the same principal vectors as tensor $\bm{U}$, one can write 
		\begin{equation}\label{Eq:7}
		\bm{C}= \displaystyle\sum_{i=1}^{3}\Lambda_i\bm{Y}_i\otimes\bm{Y}_i
		\end{equation}
		 where $\Lambda_i = \lambda_i^2$ are the principal stretches of $\bm{C}$. The same decomposition theorem can be applied to the tensors $\bm{V}$ and $\bm{B}$ \cite{holzapfel2001nonlinear}. 
		The principal invariants of tensors $\bm{C}$ and $\bm{B}$ are defined as \cite{holzapfel2001nonlinear}
		\begin{equation}\label{Eq:8}
		\begin{aligned}
		\mathcal{J}_{1\bm{C}} &= \text{tr}\bm{C} = \text{tr}\bm{B}= \mathcal{J}_{1\bm{B}}\\
		\mathcal{J}_{2\bm{C}} &=  \frac{1}{2}\bigg[(\text{tr}\bm{C})^2 - \text{tr}\bm{C}^2\bigg]=\frac{1}{2}\bigg[(\text{tr}\bm{B})^2 - \text{tr}\bm{B}^2\bigg]=\mathcal{J}_{2\bm{B}} \\
		\mathcal{J}_{3\bm{C}} &= \det\bm{C}= \det\bm{B}= \mathcal{J}_{3\bm{B}} =(\det \bm{F})^2.
		\end{aligned}
		\end{equation}
		Using $\det \bm{F} = J$(say), i.e., square root of third principal invariant $\sqrt{\mathcal{J}_{3\bm{C}}}$, the relationship between the  deformed elemental volume $\mathrm{d}v$ in terms of corresponding reference volume $\mathrm{d}V$ can be written as
		\begin{equation}\label{Eq:9}
		\mathrm{d}v = J\mathrm{d}V.
		\end{equation}
		 Deformation can be separated into dilatation (volumetric) and distortional (isochoric) forms. The latter does not imply any change in volume, i.e., $\det \bar{\bm{F}} = 1$ and thus, $\bar{\bm{F}} = J^{-1/3}\bm{F}$ (Fig. \ref{fig:fig3}). In this view and using Eq. \eqref{Eq:6}, distortional parts of the right and left Cauchy-Green tensors can be represented as
		\begin{equation}\label{Eq:10}
		\bar{\bm{C}} = \bar{\bm{F}}^T\bar{\bm{F}} = J^{-\frac{2}{3}}\bm{C},\qquad \bar{\bm{B}} = \bar{\bm{F}}\bar{\bm{F}}^T = J^{-\frac{2}{3}}\bm{B}.
		\end{equation}
		Likewise, one can also find the distortional parts of the principal invariants mentioned in Eq. \eqref{Eq:8}. 
		\subsubsection{Material Model}
		Typically, 	PDMS is widely used to fabricate microfluidic devices for various applications, in particular when deformability is important. However, many researches show that its isotropic material properties depend on various factors, e.g., thickness of the fabricated device \cite{liu2009thickness}, curing temperature \cite{liu2009influences}, ratio of mixing of pre-polymer base and cross-linking agents \cite{khanafer2009effects}, etc. Furthermore, it has been established that PDMS exhibits nonlinear characteristics, and it can be closely related to rubber-like material behaviour \cite{lotters1997mechanical}. 
		
		In literature, many realistic and reliable material descriptions/models are available to describe rubber-like materials \cite{holzapfel2001nonlinear}, which can also be realized in context of FE analysis. In general, such descriptions are represented in terms of a strain energy function $\Psi$ which can be further defined in terms of the left or right Cauchy-Green deformation tensor (Eq. \ref{Eq:6}), or the principal stretches of the deformation tensors (Eq. \ref{Eq:7}), or the invariants of deformation tensors (Eq. \ref{Eq:8}). Typically,  invariants-based models are comparatively easier to implement and  computationally less expensive in the FE setting than material definitions which utilize eigenvalues of the deformation tensors. This is because the latter involves numerically expensive operations as well as co-ordinate system transformations \cite{bacsar1998finite}. A rubber-like material exhibits different responses for dilation and distortional deformations \cite{holzapfel2001nonlinear}. Consequently, the strain energy function can be decomposed additively as
		\begin{equation}\label{Eq:11}
		\Psi = \Psi_\mathrm{dil}(J) + \Psi_\mathrm{dis}(\bar{\mathcal{J}}_{1\bm{B}},\,\bar{\mathcal{J}}_{2\bm{B}},\,\bar{\mathcal{J}}_{3\bm{B}})
		\end{equation}   
		wherein $\Psi_\mathrm{dil}(J)$ is defined by the stored energy due to a volume change $J$, $\Psi_\mathrm{dis}$ describes the elastic energy of the deviatoric deformation and $\bar{\mathcal{J}}_{1\bm{B}},\,\bar{\mathcal{J}}_{2\bm{B}},\,\bar{\mathcal{J}}_{3\bm{B}}$ are the distortional principal invariants of $\bm{B}$. Further, $\Psi_\mathrm{dil}(J)$, a strictly convex function, which reaches its unique minimum at $J = 1$, is taken as \cite{simo1992associative}
		\begin{equation}\label{Eq:12}
		\Psi_\mathrm{dil}(J) = \frac{K}{2}\bigg[\frac{J^2-1}{2} - \ln J\bigg]
		\end{equation} 
		where $K$ is the linear bulk modulus, which is also referred as the penalty parameter in context of incompressible materials wherein such materials are modeled as slightly compressible \cite{holzapfel2001nonlinear}. We use the Arruda-Boyce \cite{arruda1993three,holzapfel2001nonlinear} material definition for the deviatoric part of the strain energy $\Psi_\mathrm{dis}(\bar{\mathcal{J}}_{1\bm{B}},\,\bar{\mathcal{J}}_{2\bm{B}},\,\bar{\mathcal{J}}_{3\bm{B}})$, which is defined for rubber-like materials as
		\begin{equation}\label{Eq:13}
		\Psi_\mathrm{dis}  = a_1(\bar{\mathcal{J}}_{1\bm{B}} - 3) + a_2(\bar{\mathcal{J}}_{1\bm{B}}^2 -9) + a_3(\bar{\mathcal{J}}_{1\bm{B}}^3 -27)
		\end{equation}
		wherein $\bar{\mathcal{J}}_{1\bm{B}} = J^{-\frac{2}{3}}\mathcal{J}_{1\bm{B}}$, $a_1 = \frac{\bar{G}}{2}$, $a_2 = a_1\frac{1}{10n}$ and $a_3 = a_1\frac{11}{525n^2}$  with $\bar{G} = \frac{G}{1+\frac{3}{5n}+\frac{99}{175n^2}}$. Further, $G$ is the linear shear modulus and $n$ is the number of segments in the chain of the material molecular network structure. Note that $\Psi_\mathrm{dil}(J)=0$ and $\Psi_\mathrm{dis} = 0$ hold iff $J= 1$ and $\bar{\bm{B}} = \bm{I}$, respectively. In this paper, the value of shear modulus $G$ and bulk modulus $K$ are taken as $0.68$ MPa and $3.42$ MPa for PDMS, respectively \cite{johnston2014mechanical}. The strain energy function $\Psi$ for the employed material model with the approximated Arruda-Boyce material model (Eq. \ref{Eq:13}) and dilatation term (Eq. \ref{Eq:12}) can be written as:
		\begin{equation}\label{Eq:13.1}
		\Psi = a_1(\bar{\mathcal{J}}_{1\bm{B}} - 3) + a_2(\bar{\mathcal{J}}_{1\bm{B}}^2 -9) + a_3(\bar{\mathcal{J}}_{1\bm{B}}^3 -27) + \frac{K}{2}\bigg[\frac{J^2-1}{2} - \ln J\bigg].
		\end{equation}
		Note, the strain energy of the Arruda-Boyce material model (Eq. \ref{Eq:13}) is derived from the inverse Langevin function using a Taylor expansion \cite{arruda1993three}. It is imperative to use more Taylor series terms for better representation of the Arruda-Boyce material behavior. As per \cite{zienkiewicz2005finite}, we use the first three terms from the approximation in our FE formulation, realizing that this introduces an approximation to the ideal Arruda-Boyce material model. Nevertheless, it is found that the employed material model (Eq. \ref{Eq:13.1}) yields an adequate match of PDMS behavior for the CFCSs (Section \ref{section:4.2}), suited for use in our design optimization studies. Readers are suggested to refer Ref. \cite{BENITEZ2018153} for a detailed description and evaluation of inverse Langevin function.
		
		Following the fundamentals of non-linear continuum mechanics \cite{holzapfel2001nonlinear}, the Cauchy stress tensor $\bm{\sigma}$ and material tangent tensor $\mathbb{C}$ (see Appendix \ref{Append:A}) can be determined for the considered strain energy function $\Psi$ (Eq. \ref{Eq:13.1}). The latter is deemed necessary when using iterative/incremental solution techniques, e.g, Newton-Raphson, to solve the nonlinear problems in computational finite elasticity while the first is needed to evaluate the internal energy of the continua.
		
	\subsection{Numerical Implementation}
		 
		Numerical implementation of the topology optimization problem is relatively straightforward wherein one follows the standard procedures established in TO \cite{martin2003topology} while considering the nonlinear finite element analysis to cater to large deformation of the CFCSs. For a general discussion on nonlinear finite element analysis in detailed the readers are referred to the Refs. \cite{zienkiewicz2005finite,wriggers2008nonlinear}. 
		 
		To solve non-linear mechanical equilibrium equations stemming from large deformation, we use the Newton-Raphson (N-R) iterative technique in conjunction with the updated Lagrangian based nonlinear finite element formulation while considering a \textit{plane-stress} condition matching the planar configuration and loading conditions of the considered CFCSs. For the sake of completeness, we summarize the formulation here in general terms.
		
		To compute the state variable, i.e., displacement field $\bm{u}\in\bm{\varphi}_u$ of the mechanical equilibrium equations, one solves the following weak form \cite{zienkiewicz2005finite}
		\begin{equation} \label{Eq:14}
		\begin{split}
		\underbrace{\int_{\mathcal{B}} \nabla\bm{v}: \bm{\sigma}\,\mathrm{d} v}_{\text{internal virtual work}} \underbrace{-\int_{\partial_t\mathcal{B}} \bm{v} \cdot {\bm{t}}\,\mathrm{d}a - \int_{\mathcal{B}}\bm{v}\cdot{\rho\bm{b}}\, \mathrm{d}v}_{\text{-external virtual work}} =0 \,\,\, \forall\bm{v} \in \bm{\varphi}_v
		\end{split}
		\end{equation}
		wherein, a body $\mathcal{B}$ is in the current configuration with known volumetric body load $\bm{b}$, traction $\bm{t}$ on $\partial_t\mathcal{B}\in \partial\mathcal{B}$, and prescribed displacement boundary condition on $\partial_u\mathcal{B}\in \partial\mathcal{B}$. Note that $\partial\mathcal{B} = \partial_t\mathcal{B}\cup\partial_u\mathcal{B}$ and $\partial_t\mathcal{B}\cap\partial_u\mathcal{B}= \emptyset$. Further, $\bm{\sigma}$, $\mathrm{d}v$ and $\mathrm{d}a$ are the Cauchy stress tensor, infinitesimal volume and area, respectively. $\bm{\varphi}_u$ and $\bm{\varphi}_v$ represent the kinematically admissible displacement and its variation fields, respectively. $\nabla\bm{v}$ is the gradient of $\bm{v}$ with respect to $\bm{x}\in\mathcal{B}$.
		
		Let $\Omega_e^0$ and $\Omega_e$ be elements in the reference $\mathcal{B}_0$ and spatial $\mathcal{B}$ configurations, respectively. In the discrete setting,  within each element $\Omega_e$, the displacement field $\bm{u}_e\in\Omega_e$ and its variation $\bm{v}_e\in\Omega_e$ are approximated using the standard FE interpolation\footnote{FE interpolated quantities are indicated via superscript $h$.  Note, to write the discrete variables and (associated) field variables the normal font and italic font are used, respectively.} as
		\begin{subequations}\label{Eq:15}
		\begin{align}
		\bm{u}_e \approx \bm{u}_e^h = \mathbf{N}{\mathbf{u}}_e\\ \bm{v}_e \approx \bm{v}_e^h = \mathbf{N}{\mathbf{v}}_e
		\end{align}
		\end{subequations}  
		where $\mathbf{N} = [N_1\mathbf{I},\, N_2\mathbf{I},\,\cdots,\, N_{n_{e}}\mathbf{I}]$ are the shape functions and $\mathbf{u}_e^T = [\mathbf{u}_1^T, \,\mathbf{u}_2^T,\,\cdots,\,\mathbf{u}_{n_{e}}^T] $. Further, ${n_{e}}$ is the number of nodes in an element $\Omega_e$ ($\Omega_e^0$), $\mathbf{I}$ is the identity matrix in $\mathcal{R}^2$ and $\mathbf{u}_1, \,\mathbf{u}_2,\,\cdots,\,\mathbf{u}_{n_{e}}$ are the  displacements of nodes $1,\,2,\,\cdots,{n_{e}}$ of the element $\Omega_e$, respectively. Likewise, using the shape functions $\mathbf{N}$, the geometric fields $\bm{x}_e\in\Omega_e$ and $\bm{X}_e\in\Omega_e^0$ can be approximated. In view of Eqs. \eqref{Eq:15}a and \eqref{Eq:15}b, the weak form (Eq. \ref{Eq:14}) can be written as
		\begin{equation} \label{Eq:16}
		\mathbf{v}^T\big[\mathbf{f}_{\mathrm{int}}(\mathbf{u},\,\bm{\zeta}) - \mathbf{f}_\mathrm{ext}\big] = \mathbf{0}, \qquad \mathbf{v} \in \bm{\varphi}_v
		\end{equation}
		where $\mathbf{f}_{\mathrm{int}}(\mathbf{u},\,\bm{\zeta})$ and $\mathbf{f}_{\mathrm{ext}}$ are internal and external forces, respectively \cite{zienkiewicz2005finite}. $\mathbf{v}$ is the global vector stemming from the kinematically admissible virtual displacements of all the
		finite element nodes. Eq. \eqref{Eq:16} provides the nonlinear
		mechanical equilibrium equation with residual force $\mathbf{R}(\mathbf{u},\,\bm{\zeta})$ as
		\begin{equation} \label{Eq:17}
		\mathbf{R}(\mathbf{u},\,\bm{\zeta}) = \mathbf{f}_{\mathrm{int}}(\mathbf{u},\,\bm{\zeta}) - \mathbf{f}_\mathrm{ext} = \mathbf{0}
		\end{equation}
		  which is solved using the Newton-Raphson method at constant $\bm{\zeta}$ in the following manner
		  \begin{subequations}\label{Eq:18}
		 \begin{align}
		 \mathbf{K}^{\mathrm{g}}(\mathbf{u}_r)\Delta \mathbf{u}_{r+1} &=-\mathbf{R}(\mathbf{u}_r,\,\bm{\zeta})\\
		 \mathbf{u}_{r+1} &= \mathbf{u}_{r} + \Delta \mathbf{u}_{r+1},
		 \end{align}
		 \end{subequations}  
		 where $\mathbf{u}_r$ and $\mathbf{u}_{r+1}$ are the nodal displacements at iterations $r$ and $r+1$, respectively. $ \Delta \bm{u}_{r+1}$ is the correction in nodal displacements $\bm{u}_r$. The global tangent stiffness matrix $\mathbf{K}^{\mathrm{g}}$ is obtained via summation of internal $\mathbf{K}^{\mathrm{g}}_\mathrm{int}$ and external $\mathbf{K}^{\mathrm{g}}_\mathrm{ext}$ global stiffnesses as
		 \begin{equation}\label{Eq:19}
		 \mathbf{K}^{\mathrm{g}} = \mathbf{K}^{\mathrm{g}}_\mathrm{int} +\mathbf{K}^{\mathrm{g}}_\mathrm{ext}.
		 \end{equation}
		Herein, $\mathbf{K}^{\mathrm{g}}_\mathrm{ext}=\mathbf{0}$ as we consider cases where $\mathbf{f}_\mathrm{ext}$ is constant. $\mathbf{K}^{\mathrm{g}}_\mathrm{int}$  is obtained by assembling element stiffness matrices $\mathbf{K}^e_\mathrm{int} = \frac{\partial \mathbf{f}^e_{\mathrm{int}}}{\partial \mathbf{u}}$ (Appendix \ref{Append}). Internal element forces $\mathbf{f}^e_{\mathrm{int}}$ are evaluated as $\mathbf{f}^e_\mathrm{int} = \int_{\Omega_e}\mathbf{B}_{UL}^T\bm{\sigma} \mathrm{d}{v},$
		where $\mathbf{B}_{UL}$ is the strain-displacement matrix \cite{bathe2006finite} as per the updated Lagrangian formulation and $\bm{\sigma}$ is the Cauchy stress tensor (Appendix \ref{Append:A}).
		
		The plane-stress conditions, i.e., stresses in the direction normal to the plane of deformation $\sigma_{3i}|_{i= 1,\,2,\,3} = 0$, need to be imposed since the designed CFCSs are planar. It is also required to account for the change in dimension pertaining to normal to the plane of deformation. This is done by representing the deformation gradient as
		\begin{equation}\label{Eq:20}
		\bm{F}= \begin{bmatrix}
		F_{11} & F_{12} &  0\\
		F_{21} & F_{22} &  0\\
		0 & 0 &  F_{33}\\
		\end{bmatrix} 
		\end{equation} 
		wherein $F_{33}$ is obtained from material constitution by following the local iteration form approach \cite{klinkel2002using,zienkiewicz2005finite} wherein $\sigma_{3i}|_{i= 1,\,2,\,3} = 0$ are enforced. In the spatial configuration, the current volume is evaluated as
		\begin{equation}\label{Eq:21}
		J = \det \bm{F} = (F_{11}F_{22} - F_{12}F_{21})F_{33}
		\end{equation} 
		and the current thickness $t$ is calculated as
		\begin{equation}\label{Eq:22}
		t = HF_{33}
		\end{equation}
		where $H$ is the thickness in the reference configuration.
	
		\section{Objective Formulation and Sensitivity Analysis}\label{section:3}
\begin{figure}[h!]
			\begin{subfigure}[t]{0.450\textwidth}
				\centering
				\includegraphics[scale=1]{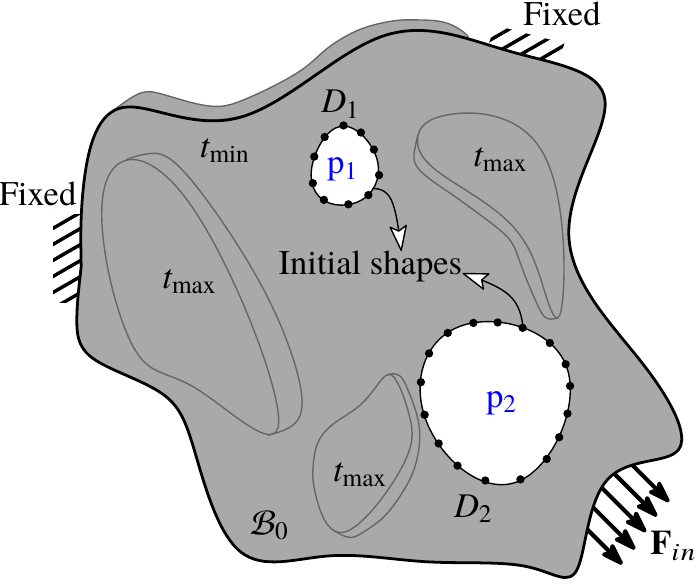}
				\caption{Initial configuration}
				\label{fig:fig4a}
			\end{subfigure}%
			~ \qquad
			\begin{subfigure}[t]{0.45\textwidth}
				\centering
				\includegraphics[scale=1]{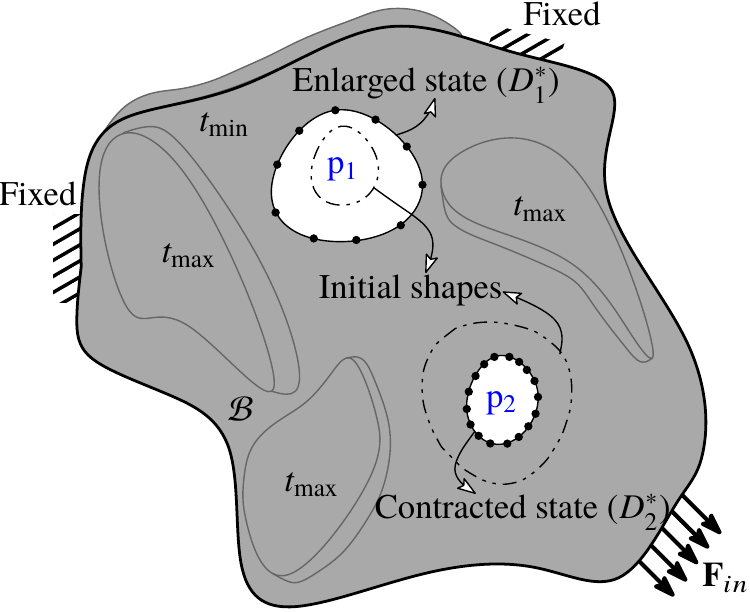}
				\caption{Deformed configuration}
				\label{fig:fig4b}
			\end{subfigure}
			\caption{The initial and deformed configurations of a CFCS containing two functional pores p$_1$ and p$_2$ (output locations) with fixed boundary conditions are depicted. As the actuating force $\mathbf{F}_{in}$ acts, it is desired that pores p$_1$ and p$_2$ eventually are enlarged and contracted, respectively. For contracting and enlarging, the boundary of the pores should move towards and apart from each other and thereby, reducing and increasing the respective hydraulic diameters.  In FE setting, say a pore is constituted via $k = 1,\,2,\,\cdots,\,nbn$ nodes. To achieve the respective states (solid lines in Fig. \subref{fig:fig4b}) in response to the external loading, we prescribe different hydraulic diameter targets $D_i^*$ to the pores and extremize the objective $f_0$ (Eq. \ref{Eq:24}).}\label{fig:fig4}
		\end{figure}

		As aforementioned, the output responses of the synthesized CFCSs can be characterized via either enlarging or contracting or various combinations of both enlarging and/or contracting of the functional pores. These pores with known initial positions, shapes and sizes, are the integral part of the CFCSs, and they are predefined within the given design domain $\mathcal{B}_0$. Particularly, embedded pores of a CFCS can eventually attain either of the two states with respect to their initial configurations: (i) enlarged state and (ii) contracted state, when external loads are applied. Herein, hydraulic diameter $D_i$ \cite{hetsroni2005fluid} is employed to indicate the state of the $i^\mathrm{th}$ functional pore and its effect on the flow resistance, i.e., an increment and a reduction in $D_i$ imply that the $i^\mathrm{th}$ pore gets enlarged and contracted, respectively (Fig. \ref{fig:fig4}). The hydraulic diameter $D_i$ is defined as
		\begin{equation}\label{Eq:23}
		D_i = \frac{4 A_i}{P_i},
		\end{equation} 
		with $ A_i$ and $P_i$ are the cross-sectional area and perimeter of the pore, respectively. Therefore, it is natural to use hydraulic diameter to define the objective function $f_0$ herein. The objective function is defined as
		\begin{equation}\label{Eq:24}
		f_0 = \displaystyle\sum_{i=1}^{nfp}\bigg(D_i^* - D_i\bigg)^2
		\end{equation}
		where $D_i^*$ is the targeted/desired hydraulic diameter of the $i^\mathrm{th}$ functional pore and $nfp$ is the total number of pores in the CFCS. In the discrete setting, a pore is defined by nodes situated on its boundary. Say, the $i^\mathrm{th}$ pore is defined by $k = 1,\,2,\,3,\cdots,nbn$ boundary nodes. These nodes are arranged in clockwise with respect to either $x-$ or $y-$coordinates and stacked into array $\mathbf{NBN}$ with their respective nodal coordinates. The arrangement is done in this manner so that the area of the pore can be evaluated appropriately. The updated positions $\mathbf{xn}$ of the nodes  are determined using the reference nodal positions $\mathbf{Xn}$ and the global displacements vector $\mathbf{u}$ obtained via FE solution. Therefore, the nodal coordinates $\mathbf{xnp}_{i}$ of the boundary nodes  of the pores are extracted.
		
		 These steps are followed to evaluate the area of the pore:
		\begin{enumerate}
			\item The coordinates of the center of the pore are determined.
			\item The pore is divided into $nbn$ triangles by considering the center point as one of their vertices. 
			\item The area $ A_i$ of the pore is determined by finding and summing the areas of $nbn$ triangles.
		\end{enumerate}
	The perimeter $P_i$ of the pore is found by summation of the lengths of each constituting side of the pore. Thereafter, using Eq. \eqref{Eq:23}, the respective hydraulic diameter is evaluated and hence, the objective (Eq. \ref{Eq:24}) is determined.
					
		 We use the \textit{adjoint variable method} \cite{ryu1985structural} which requires the solution of Lagrange multipliers associated with the mechanical equilibrium equations, to evaluate the sensitivities with respect to the design variables. In general, the adjoint variable method is suitable where the number of design variables exceeds that of state-dependent responses \cite{haug1979applied}. The augmented performance function $\bm{\Phi}(\mathbf{u},\,\bm{\zeta})$ is defined using the objective function (Eq. \ref{Eq:24}) and mechanical equilibrium equation (Eq. \ref{Eq:17}) as: 
		 \begin{equation}\label{Eq:25}
		 \bm{\Phi}(\mathbf{u},\,\bm{\zeta}) = f_0(\mathbf{u}) + \bm{\Lambda}^T \mathbf{R}(\mathbf{u},\,\bm{\zeta})
		 \end{equation}
		 
		 To evaluate the sensitivities, Eq. \eqref{Eq:25} is differentiated with respect to the design variables, which yields
		
		\begin{equation}\label{Eq:26}
		\frac{d\bm{\Phi}(\mathbf{u},\,\bm{\zeta})}{d \bm{\zeta}} = \underbrace{\bigg[\frac{\partial f_0(\mathbf{u})}{\partial\mathbf{u}}\bigg|_{\bm{\zeta}} + \bm{\Lambda}^\mathrm{T} \frac{\partial \mathbf{R}(\mathbf{u},\,\bm{\zeta})}{\partial\mathbf{u}}\bigg|_{\bm{\zeta}}\bigg]}_{\text{Term 1}} \frac{\partial\mathbf{u}}{\partial\bm{\zeta}} + \frac{\partial f_0(\mathbf{u})}{\partial\bm{\zeta}}\bigg|_{\mathbf{u}} + \bm{\Lambda}^\mathrm{T}\frac{\partial\mathbf{R}(\mathbf{u},\,\bm{\zeta})}{\partial\bm{\zeta}}\bigg|_{\mathbf{u}}
		\end{equation}
		
		where $\bm{\Lambda}$ is the Lagrange multiplier vector and one evaluates the quantity $\big(\frac{\partial \alpha\,}{\partial \beta\,}\big)\big|_{\gamma}$ as a partial derivative of $\alpha$ with respect to $\beta$ keeping variable $\gamma$ constant.  $\bm{\Lambda}$ is chosen such that Term 1 vanishes \cite{martin2003topology}. Therefore, in view of Eq. (\ref{Eq:19}), Term 1 becomes\footnote{For readability, we use $\mathbf{K}^\mathrm{g}$ as $\mathbf{K}_\mathrm{g}$ wherever it is required and vice versa.}
		\begin{equation}\label{Eq:27}
		\begin{split}
		\bm{\Lambda}^T &= -\bigg(\frac{\partial f_0(\mathbf{u})}{\partial\mathbf{u}}\Bigg|_{\bm{\zeta}}\bigg)\mathbf{K}_\mathrm{g}^{-1} \\
		&= 8\displaystyle\sum_{ifp=1}^{nfp}\left(\left(D_i^* - D_i\right)\left(\frac{P_i\mathbf{A}_{i,\mathbf{u}} - A_i\mathbf{P}_{i,\mathbf{u}}}{P_i^2}\right)\right)\mathbf{K}_\mathrm{g}^{-1}
		\end{split}
		\end{equation}
		
		where $\mathbf{K}_\mathrm{g}$ is defined in Eq. (\ref{Eq:19}) and the terms $\mathbf{A}_{i,\mathbf{u}}$ and $\mathbf{P}_{i,\mathbf{u}}$ indicate derivatives of the area $A_i$ and perimeter $P_i$ of the pore with respect to displacement vector $\mathbf{u}$. By combining Eq. \eqref{Eq:26} and Eq. \eqref{Eq:27}, the objective sensitivity is found as:
		 \begin{equation}\label{Eq:28}
		 \frac{df_0(\mathbf{u})}{d \bm{\zeta}} =  \bm{\Lambda}^\mathrm{T}\frac{\partial\mathbf{R}(\mathbf{u},\,\bm{\zeta})}{\partial\bm{\zeta}}\bigg|_{\mathbf{u}}.
		 \end{equation}
		The term $\frac{\partial\mathbf{R}(\mathbf{u},\,\bm{\zeta})}{\partial\bm{\zeta}}\bigg|_{\mathbf{u}}$ is determined from  Eq. (\ref{Eq:17}) using the expression for $\mathbf{f}_\mathrm{int}$ (see Appendix \ref{Append}).
		\section{Results and Discussion}\label{section:4}
		This section presents the results of the proposed CFCS design optimization approach, to control pores in a PDMS sheet in response to external stretching deformation. Gradual enlarging and/or contracting of embedded functional  pores provides regulated fluid flow which is here defined in terms of their hydraulic diameters.
	
		\begin{figure}[H]
			\centering
			\begin{subfigure}{0.3\textwidth}
				\includegraphics[scale =0.75]{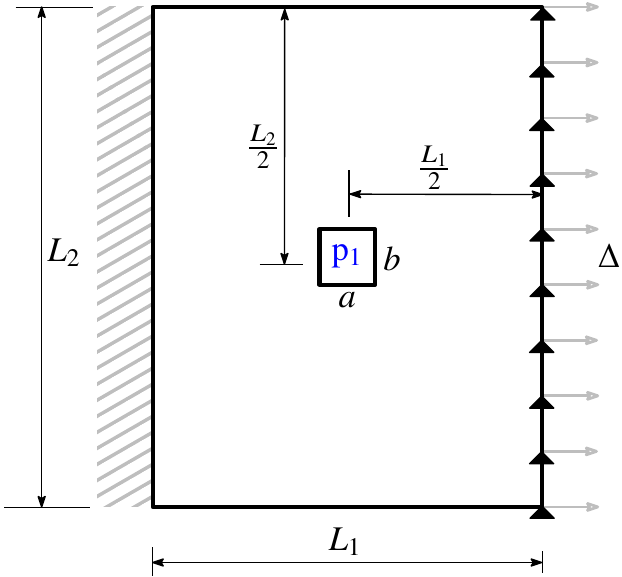}
				\caption{Design case 1}
				\label{fig:fig5a}
			\end{subfigure}
			\quad
			\begin{subfigure}{0.3\textwidth}
				\includegraphics[scale =0.75]{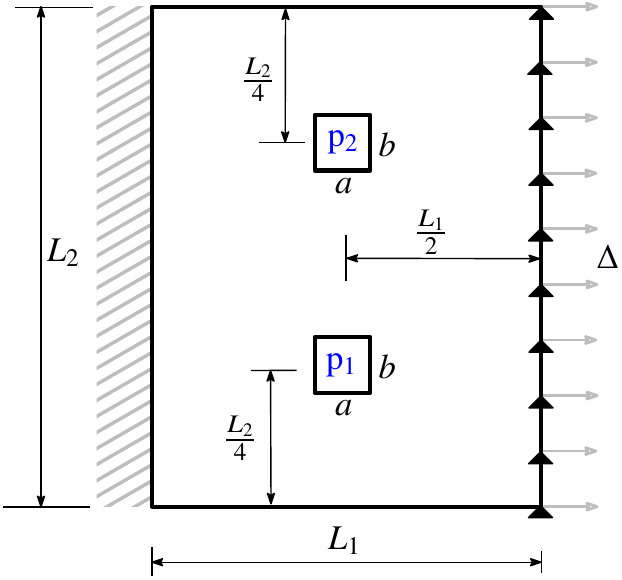}
				\caption{Design case 2}
				\label{fig:fig5b}
			\end{subfigure}
			\quad
			\begin{subfigure}{0.3\textwidth}
				\includegraphics[scale =0.75]{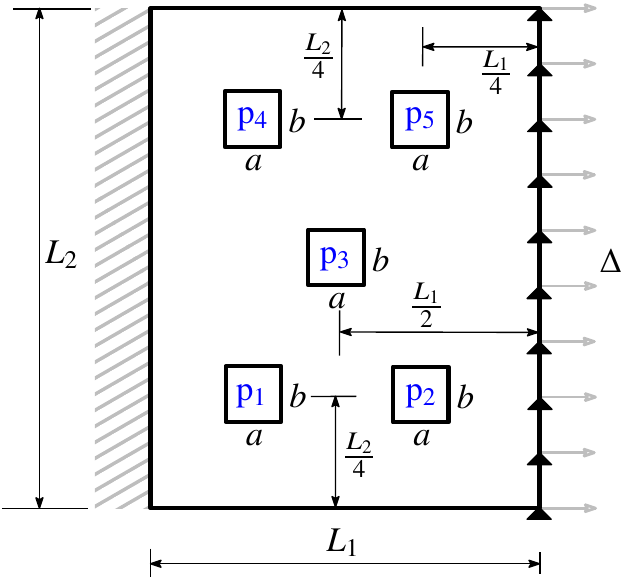}
				\caption{Design case 3}
				\label{fig:fig5c}
			\end{subfigure}
			\caption{Design specifications for CFCS I (Fig. \subref{fig:fig5a}), CFCS II (Fig. \subref{fig:fig5a}), CFCS III (Fig. \subref{fig:fig5b}) and CFCS IV (Fig. \subref{fig:fig5c}) are shown.  \textbf{Key:} $L_1 = 10$ mm, $L_2 = 16$ mm, $a=0.5$ mm, $b= 0.5$ mm and $\Delta = 10$ mm is the specified displacement.}\label{fig:fig5}
		\end{figure}
	\begin{table}[h!]
		\centering
		\begin{tabular}{ c|c|c|c }
			\hline
			\hline
			\textbf{Design cases} & \textbf{Considered pores} & \textbf{Desired function(s)} & \textbf{Center position of the pore(s)} \rule{0pt}{2ex}  \\ \hline
		    CFCS I & p$_1$ & Enlarging & ($\frac{L_1}{2},\,\frac{L_2}{2}$)  \rule{0pt}{3ex}\\ \hline
			CFCS II & p$_1$ & Contracting & ($\frac{L_1}{2},\,\frac{L_2}{2}$)\rule{0pt}{3ex}\\ \hline
			\multirow{2}{*}{CFCS III} & p$_1$ & Contracting & ($\frac{L_1}{2},\,\frac{L_2}{4}$) \rule{0pt}{3ex}\\
			& p$_2$ & Enlarging& ($\frac{L_1}{2},\,\frac{3L_2}{4}$) \rule{0pt}{3ex}\\ \hline
			\multirow{5}{*}{CFCS IV} & p$_1$ & Contracting & ($\frac{L_1}{4},\,\frac{L_2}{4}$)\rule{0pt}{3ex}\\
			& p$_2$ & Enlarging& ($\frac{3L_1}{4},\,\frac{L_2}{4}$) \rule{0pt}{3ex}\\
			& p$_3$ & Contracting& ($\frac{L_1}{2},\,\frac{L_2}{2}$)\rule{0pt}{3ex}\\
			& p$_4$ & Enlarging & ($\frac{L_1}{4},\,\frac{3L_2}{4}$)\rule{0pt}{3ex}\\ 
			& p$_5$ & Contracting & ($\frac{3L_1}{4},\,\frac{3L_2}{4}$)\rule{0pt}{3ex}\\\hline
			\hline
		\end{tabular}
		\caption{The desired function(s) of various functional pores of the CFCSs}\label{Table:Output}
	\end{table}
		\subsection{Numerical Examples}
		To show the versatility of the presented approach, four optimized CFCFs are presented with various desired output characteristics (Table \ref{Table:Output}) using the different design specifications (Fig. \ref{fig:fig5}). Prototypes of CFCS I and CFCS II have been fabricated using PDMS material and their measured performances are also compared with respective numerical models. 
		\subsubsection{Design specifications}
		The design specifications for CFCS I (Fig. \ref{fig:fig5a}), CFCS II (Fig. \ref{fig:fig5a}), CFCS III  (Fig. \ref{fig:fig5b}) and CFCS IV (Fig. \ref{fig:fig5c}) are shown in Fig. \ref{fig:fig5}. The length and width of the design domains are set to be $L_1= 10$ mm and $L_2 = 16$ mm, respectively. This is relatively large compared to the targeted microfluidic applications, but for ease of manufacturing and testing (Section \ref{FT}) this dimension was chosen. For each design domain, the left edge is fixed. The right edge of each design domain  is used to apply uniform stretching $\Delta$ in the positive horizontal direction while keeping its displacements in vertical direction unaltered. CFCS I and CFCS II have  one predefined embedded pore p$_1$ of size $a\times b = 0.5\times 0.5$ mm$^2$ in the center of their design domains (Fig. \ref{fig:fig5a}). The size of the pore is chosen considering the manufacturing limits of the used fabricating technique (Section \ref{FT}). Two functional pores p$_1$ and p$_2$, each having size of $a\times b$ mm$^2$ (Table \ref{Table:Output}) exist within the design domain of CFCS III (Fig. \ref{fig:fig5b}). Likewise, the design domain of CFCS IV (Fig. \ref{fig:fig5c}) contains five functional pores $p_i|_{i = 1,\,2,\,3,\,4,\,5}$ each having the same dimension $a\times b$ mm$^2$ (Table \ref{Table:Output}).

		\subsubsection{Optimized CFCS continua}
		The design domains (Fig. \ref{fig:fig5}) are first discretized using $Nelx\times Nely\, (= 80\times 128)$ quadratic finite elements using bilinear shape functions. $Nelx$ and $Nely$ are the number of quadratic elements in horizontal and vertical directions, respectively. The elements which lie within the functional pore(s) (Table \ref{Table:Output}) are subsequently removed. Thereafter, the element connectivity matrix and nodal numbers with corresponding coordinates are updated and are further used in the nonlinear FE analysis. To solve Eq. (\ref{Eq:17}), we use up to 10 Newton-Raphson iterations. For the optimization, $50$ iterations of MMA are performed. The target hydraulic diameters $D^*$ are set to $0$ mm and $2$ mm for contracting and enlarging, respectively. Following the limitations of the fabrication technique, the ratio $\frac{t_{\max}}{t_{\min}}$ is set to $4:1$, wherein $t_{\max}$ and $t_{\min}$ are taken as 2 mm and 0.5 mm, respectively. The penalty parameter $p$ is set to 1 (Eq. \ref{Eq:1}). The density filter radius $r_{\min}$ is set to $2\min({\frac{L_1}{Nelx},\,\frac{L_2}{Nely}})$ mm. The maximum volume limit is set to 60\% for each problem, however one can also relax this constraint and permit the optimizer to select the required amount of material. 
		
\begin{figure}[h!]
	\begin{subfigure}{0.275\textwidth}
		\centering
		\includegraphics[scale =0.5]{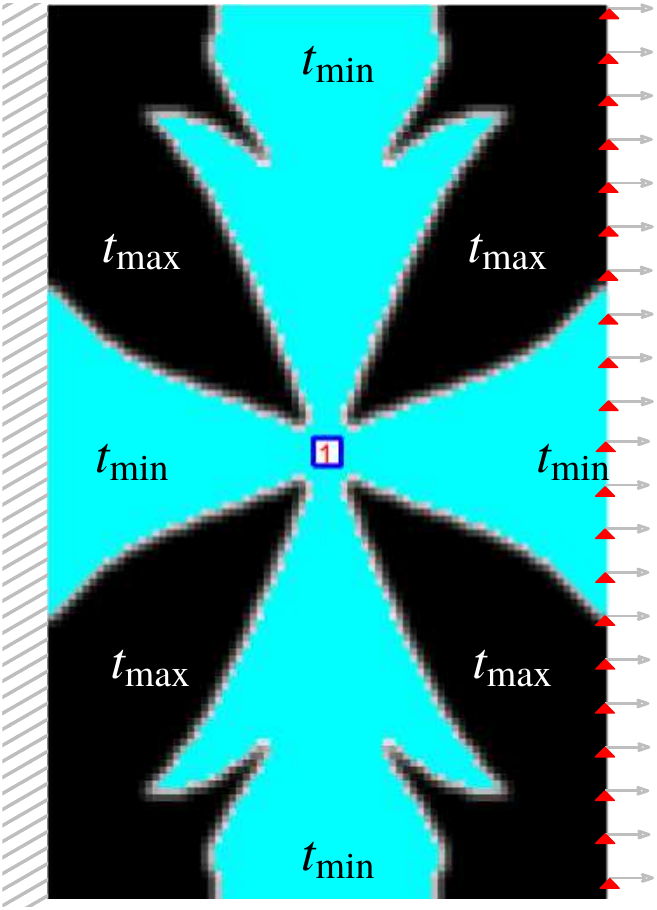}
	    \caption{CFCS I: Undeformed}
		\label{fig:fig6a}
	\end{subfigure}
	\begin{subfigure}{0.4\textwidth}
		\centering
		\includegraphics[scale =0.5]{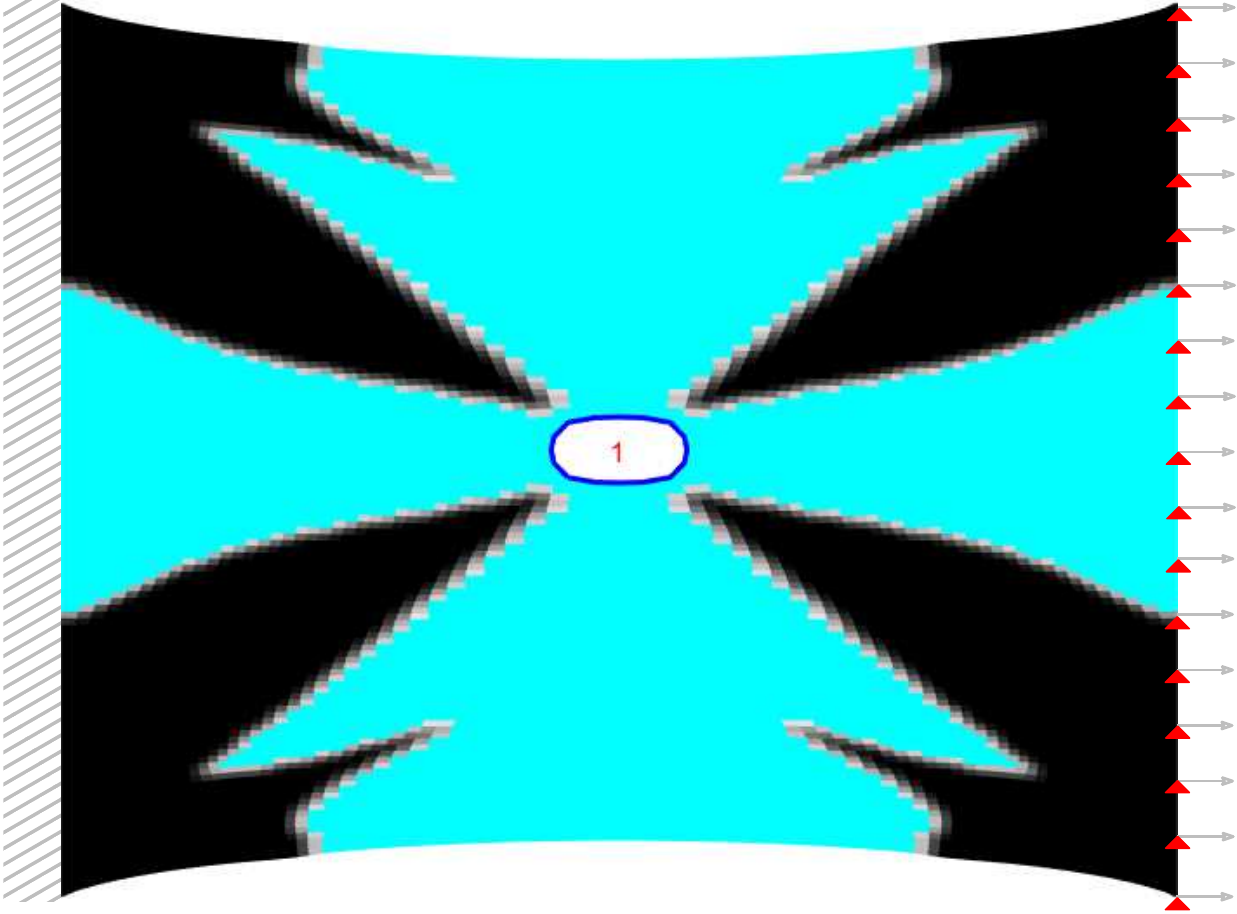}
		\caption{CFCS I: Deformed}
		\label{fig:fig6b}
	\end{subfigure}
	\begin{subfigure}{0.275\textwidth}
		\centering
		\includegraphics[scale =0.415]{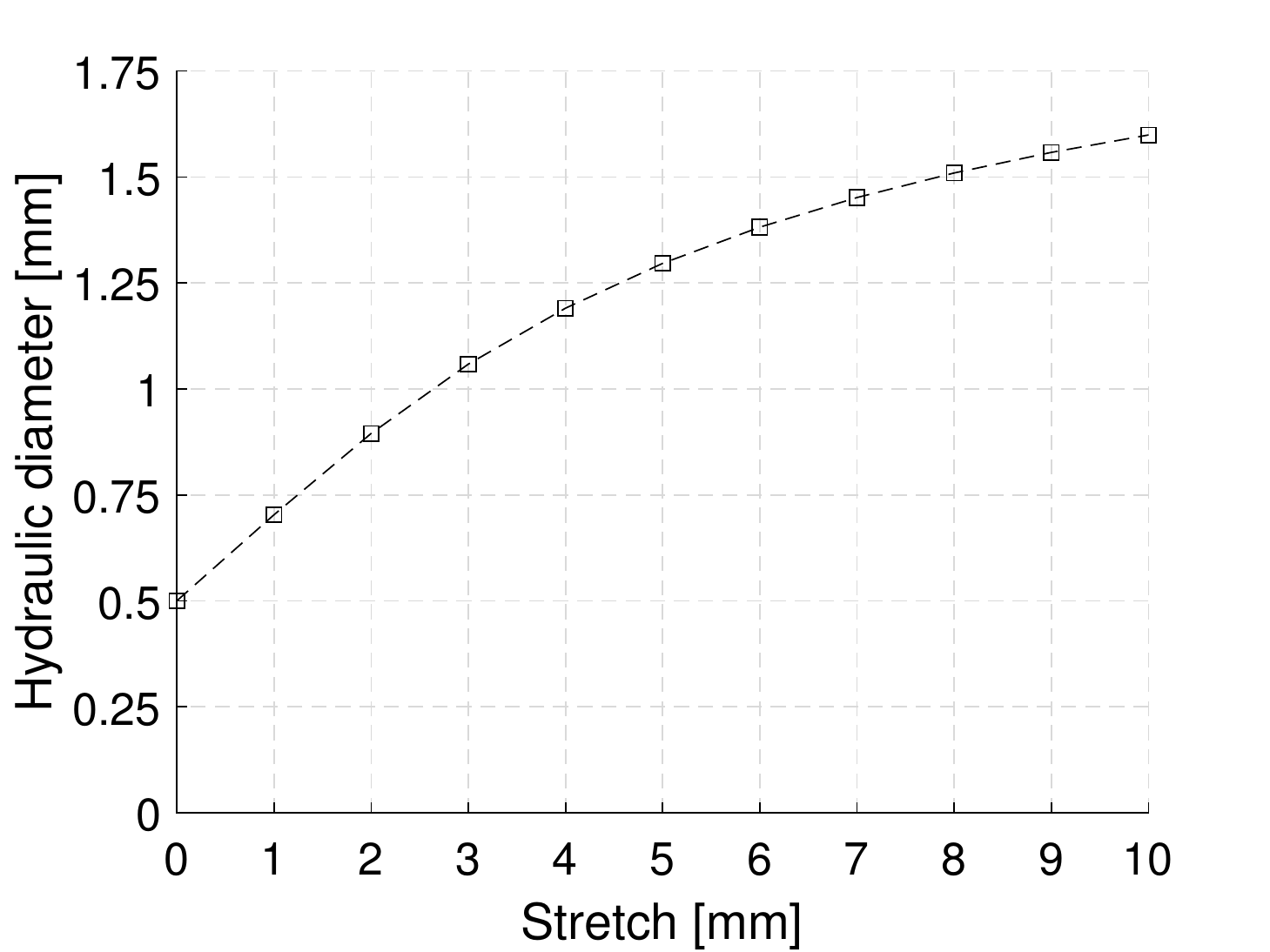}
		\caption{CFCS I: Stretch Vs HD}
		\label{fig:fig6c}
	\end{subfigure}
	\begin{subfigure}{0.25\textwidth}
		\centering
	\includegraphics[scale =0.5]{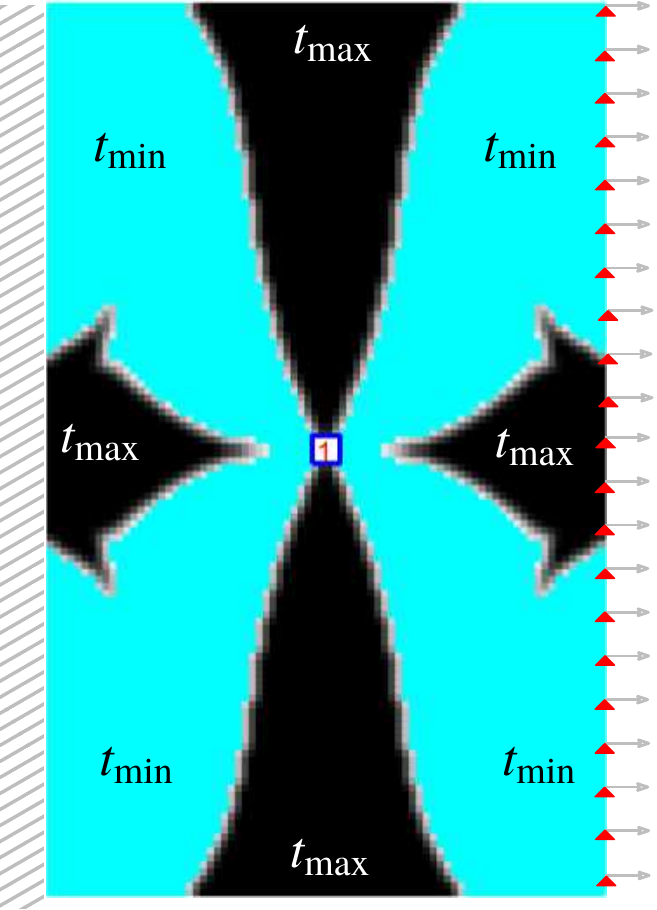}
	\caption{CFCS II: Undeformed}
	\label{fig:fig6d}
\end{subfigure}
\begin{subfigure}{0.4\textwidth}
	\centering
	\includegraphics[scale =0.5]{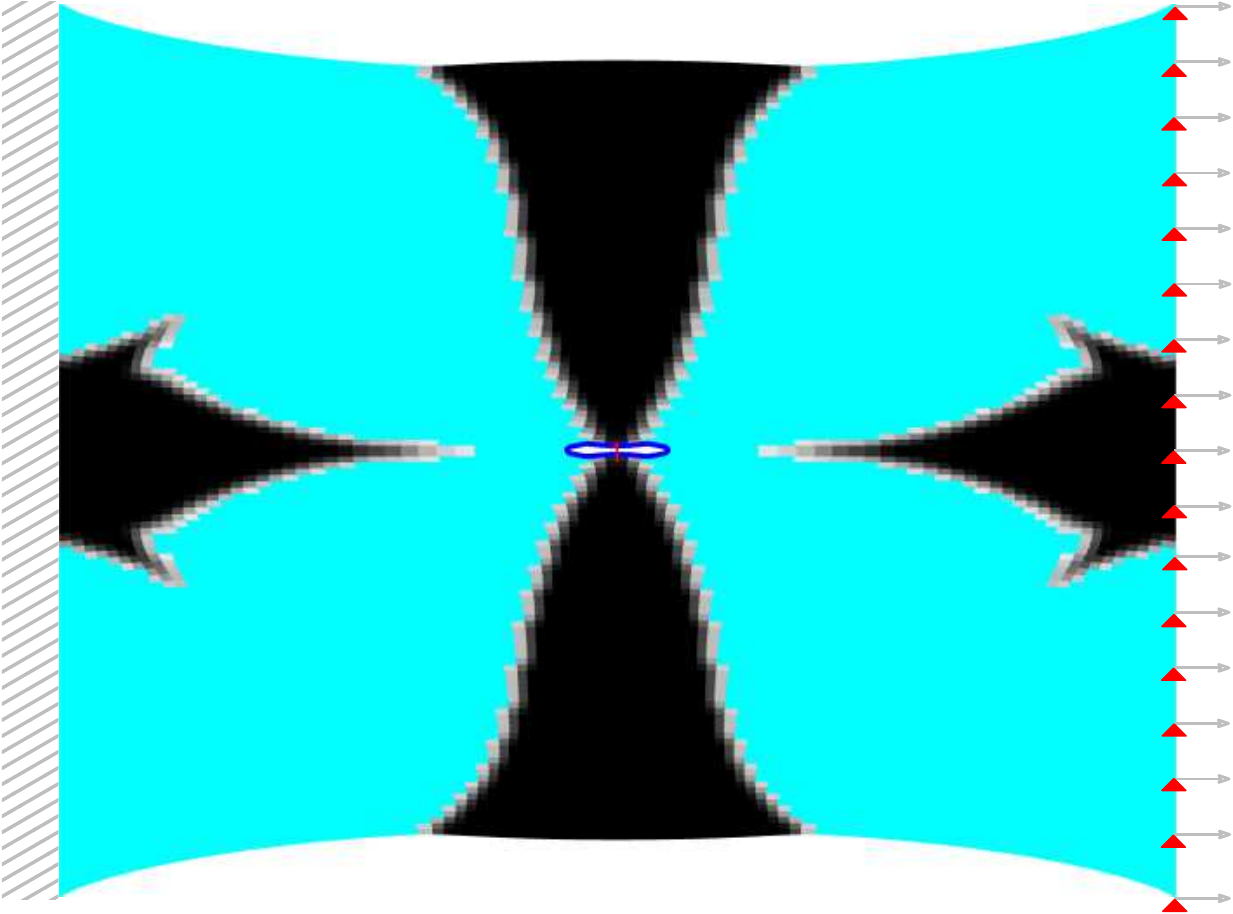}
	\caption{CFCS II: Deformed}
	\label{fig:fig6e}
\end{subfigure}
\begin{subfigure}{0.275\textwidth}
	\centering
	\includegraphics[scale =0.415]{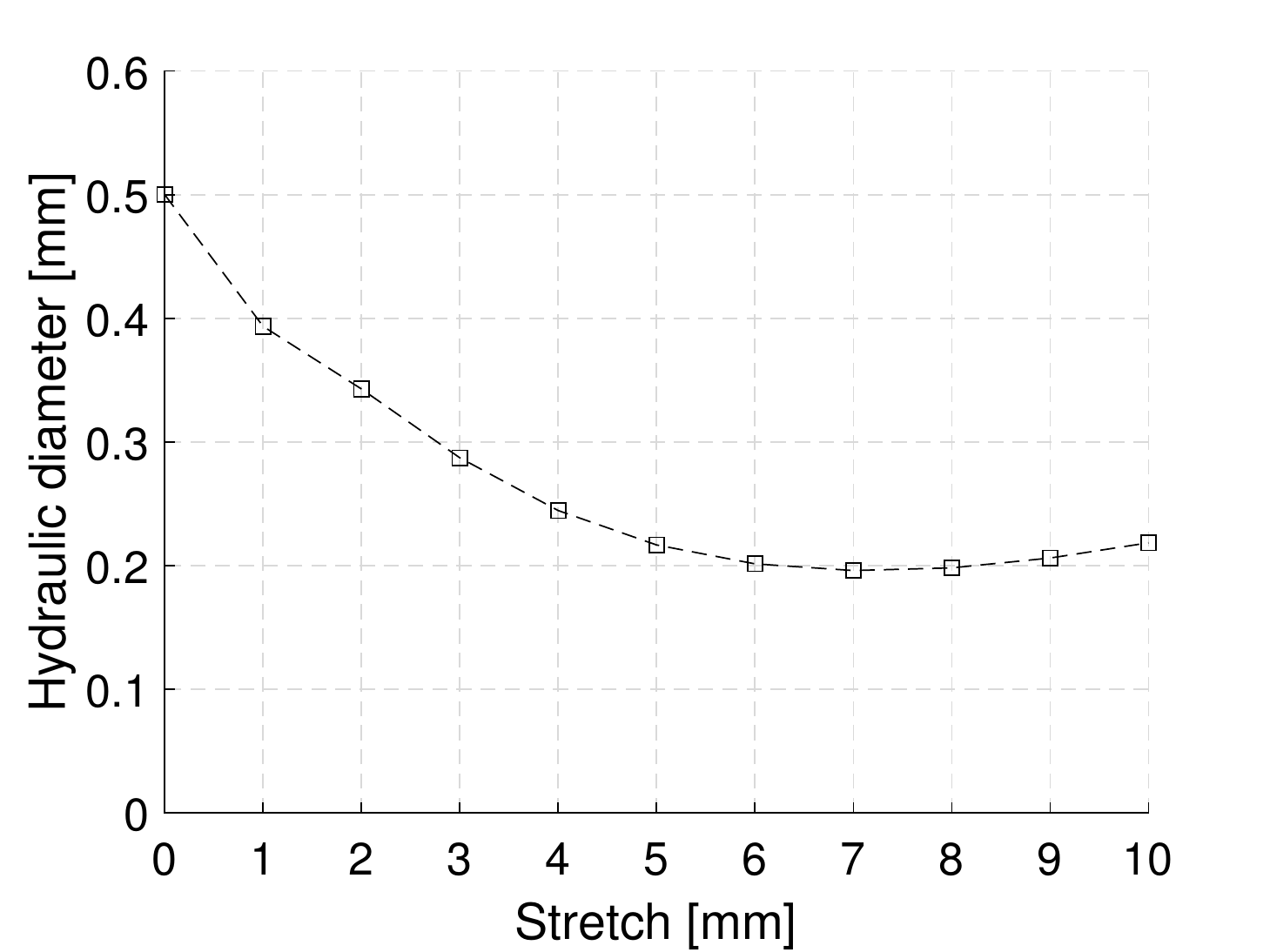}
	\caption{CFCS II: Stretch Vs HD}
	\label{fig:fig6f}
\end{subfigure}
	\begin{subfigure}{0.25\textwidth}
		\centering
	\includegraphics[scale =0.5]{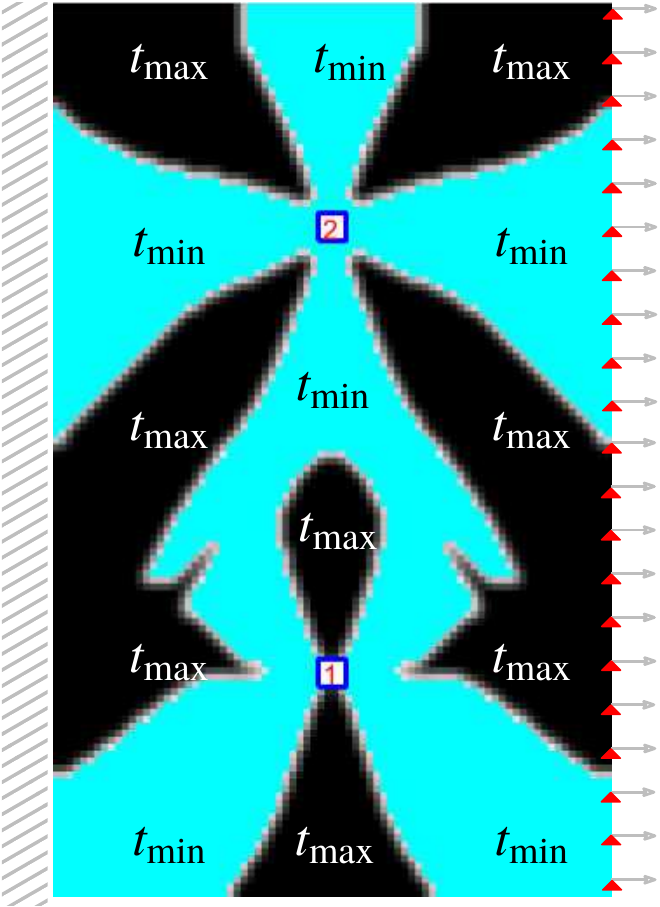}
	\caption{CFCS III: Undeformed}
	\label{fig:fig6g}
\end{subfigure}
\begin{subfigure}{0.4\textwidth}
	\centering
	\includegraphics[scale =0.5]{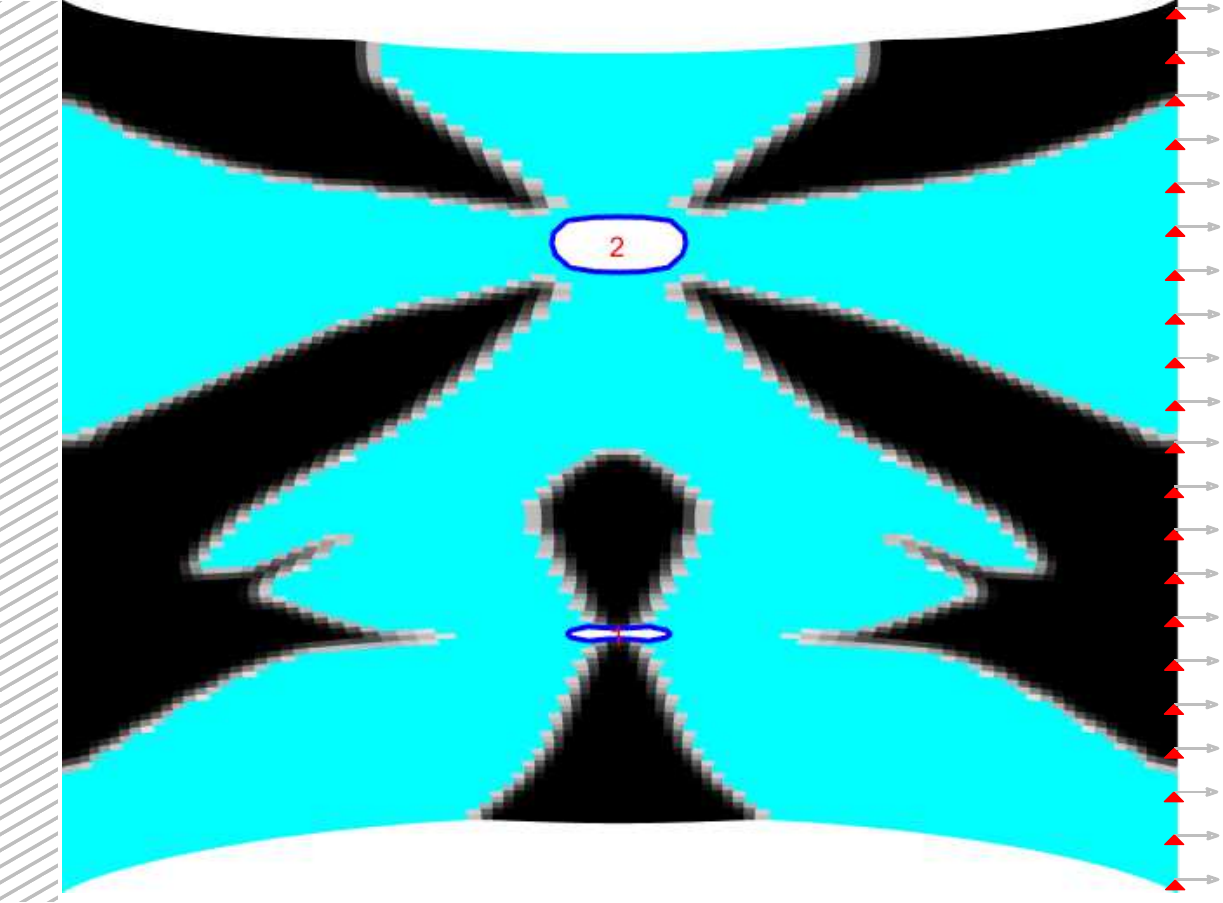}
	\caption{CFCS III: Deformed}
	\label{fig:fig6h}
\end{subfigure}
\begin{subfigure}{0.3\textwidth}
	\centering
	\includegraphics[scale =0.415]{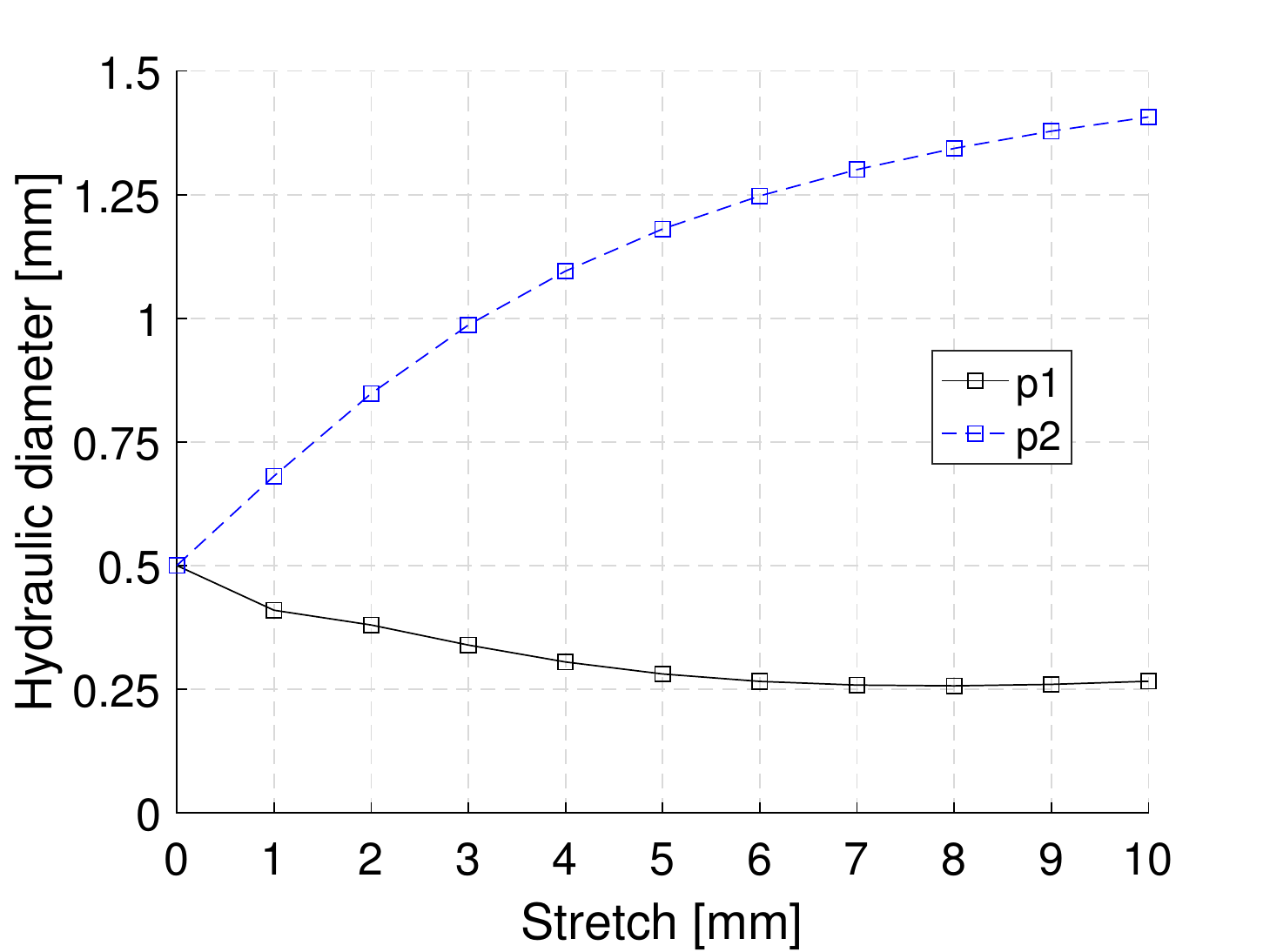}
	\caption{CFCS III: Stretch Vs HD}
	\label{fig:fig6i}
\end{subfigure}
	\begin{subfigure}{0.275\textwidth}
		\centering
	\includegraphics[scale =0.5]{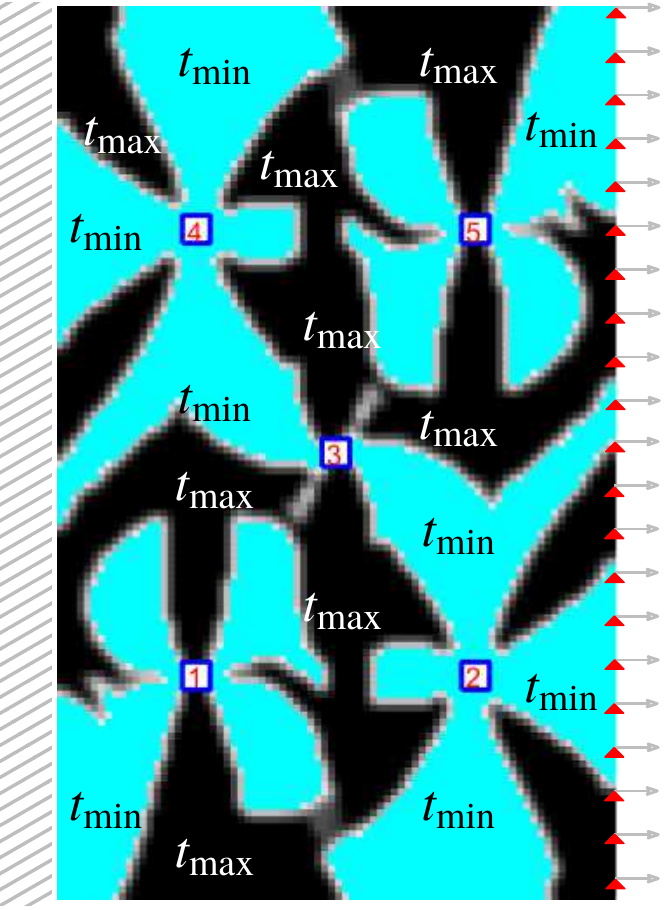}
	\caption{CFCS IV: Undeformed}
	\label{fig:fig6j}
\end{subfigure}
\begin{subfigure}{0.4\textwidth}
	\centering
	\includegraphics[scale =0.5]{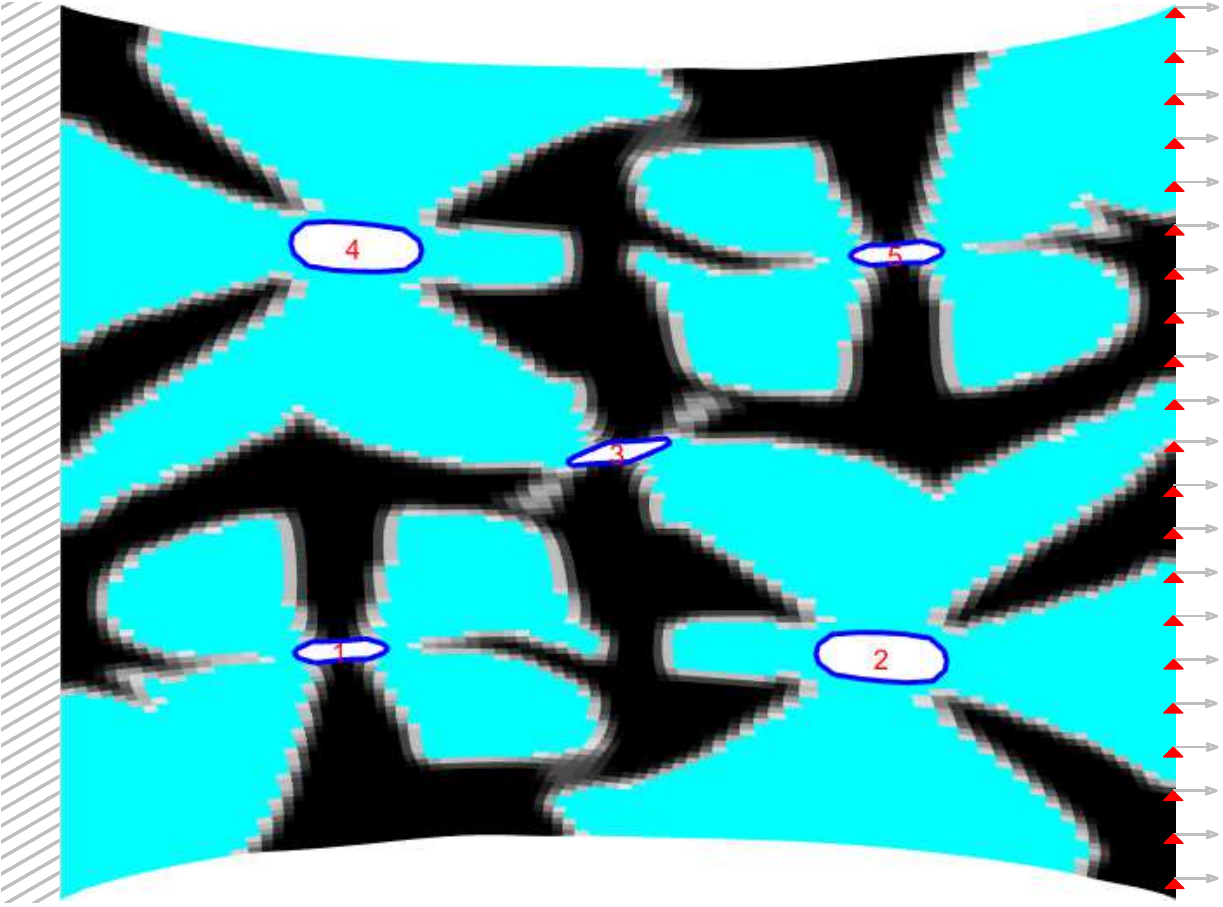}
	\caption{CFCS IV: Deformed}
	\label{fig:fig6k}
\end{subfigure}
\begin{subfigure}{0.275\textwidth}
	\centering
	\includegraphics[scale =0.415]{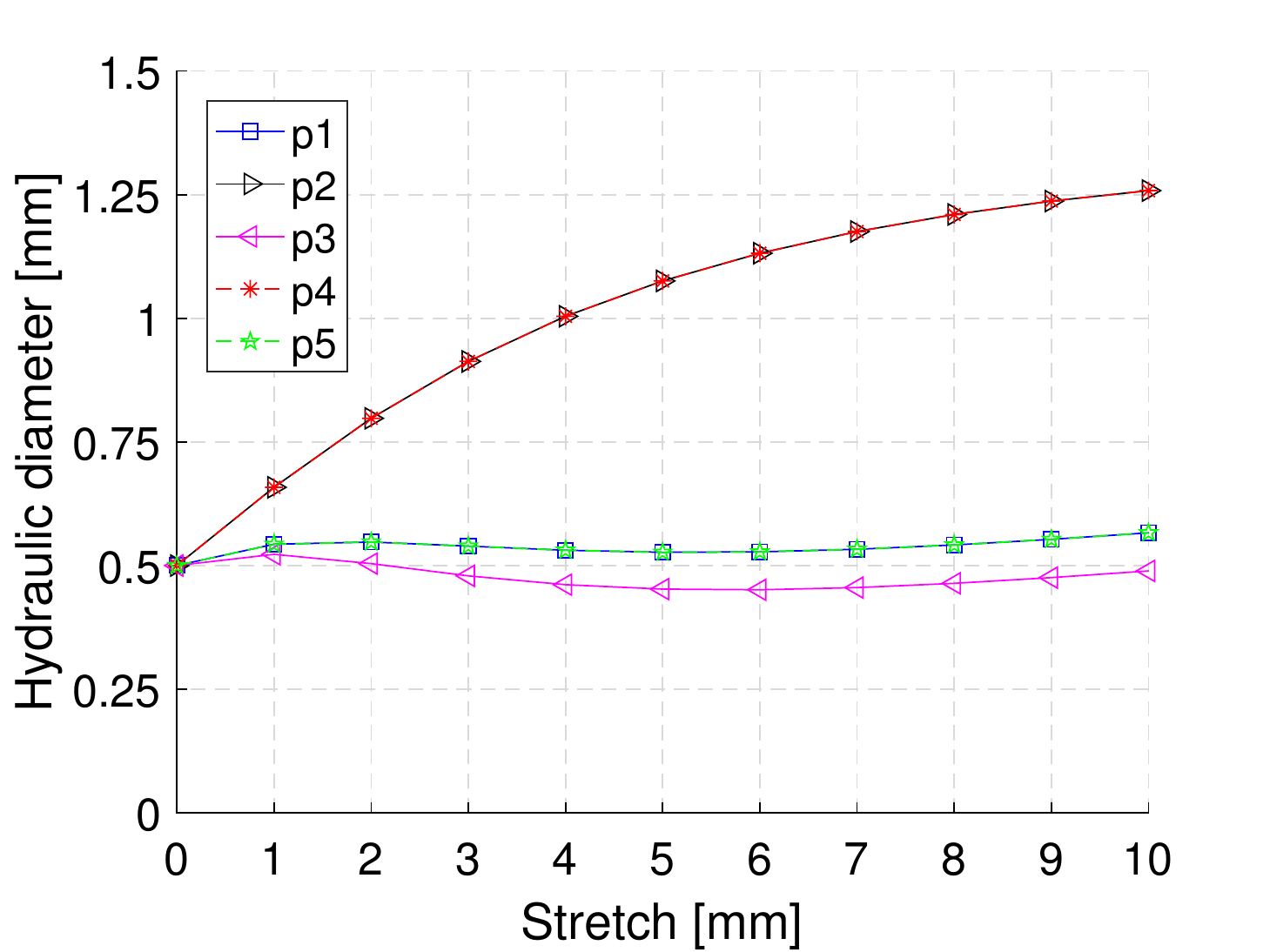}
	\caption{CFCS IV: Stretch Vs HD}
	\label{fig:fig6l}
\end{subfigure}
	\caption{The undeformed, deformed and hydraulic diameter (HD) plots versus the external response (stretch) for all optimized CFCSs. Regions containing maximum thickness $t_{\max}$ and minimum thickness $t_{\min}$ are indicated by black and cyan color. Number in red in first and second columns indicate(s) the pore identification number. \textbf{Key}: $t_{\max}= 2$ mm and $t_{\min}= 0.5$ mm.}\label{fig:fig6}
\end{figure}

		The final solutions of the four CFCS continua for the different desired pore functionalities (Table \ref{Table:Output}) are depicted in the first column of Fig. \ref{fig:fig6}.  Regions in black and cyan suggested by the optimization within the continua indicate maximum $t_{\max}$ and minimum $t_{\min}$ thicknesses. For the  objective (Eq. \ref{Eq:24}) used herein with the considered design cases (Table \ref{Table:Output}), the penalty parameter $p =1$ gives close to binary $t_{\max}$ and $t_{\min}$ thicknesses for the optimized CFCSs (Fig. \ref{fig:fig6}). However, some elements with gray color can still be observed (Fig. \ref{fig:fig6}), which indicates intermediate thicknesses for the regions associating such elements. This is a consequence of the application of density filtering, but does not prevent interpretation of the designs in the post-processing. To obtain the final design of CFCS I (Fig. \ref{fig:fig6a}), CFCS II (Fig. \ref{fig:fig6d}), CFCS III (Fig. \ref{fig:fig6g}) and CFCS IV (Fig. \ref{fig:fig6i}) continua, the optimizer uses $52.1$\%, $33.5$\%, $45.3$\% and $44.3$\% of maximum volume, respectively. The smooth convergence histories for CFCSs are depicted in Fig. \ref{fig:fig12} (Appendix \ref{Append:B}). The second column of Fig. \ref{fig:fig6} shows the deformed configurations of the CFCSs (to scale) when a stretching displacement of 10 mm is applied in the positive horizontal direction, resulting in a global strain of 100\%. The variations of hydraulic diameter(s) with respect to the input displacement for the corresponding functional pore(s) of CFCSs are depicted in the third column of Fig. \ref{fig:fig6}. Note that one can also regard input loads as  design variables \cite{kumar2016synthesis}  which is not considered herein. For the applications targeted in this study a prescribed displacement is considered more relevant.
		
		The functional pores of CFCS I and CFCS II are designed for different output responses (Table \ref{Table:Output}). In response to  external stretching, the function pore of CFCS I gradually enlarges, i.e., its corresponding hydraulic diameter increases (Fig. \ref{fig:fig6c}) and thus, permits higher fluid/nutrient/drug transport. In contrast, the embedded functional pore of the CFCS II eventually contracts  (Fig. \ref{fig:fig6f}) and thereby reduces fluid/nutrient/drug transport with the actuation. CFCS III contains two functional pores which behave differently as specified (Table \ref{Table:Output}) in response to the external stretching (Fig. \ref{fig:fig6}). The pore p$_1$ contracts eventually whereas pore p$_2$ enlarges (Fig. \ref{fig:fig6}). 

\begin{figure}[h!]
			\begin{subfigure}[t]{0.30\textwidth}
				\centering
				\includegraphics[scale=0.35]{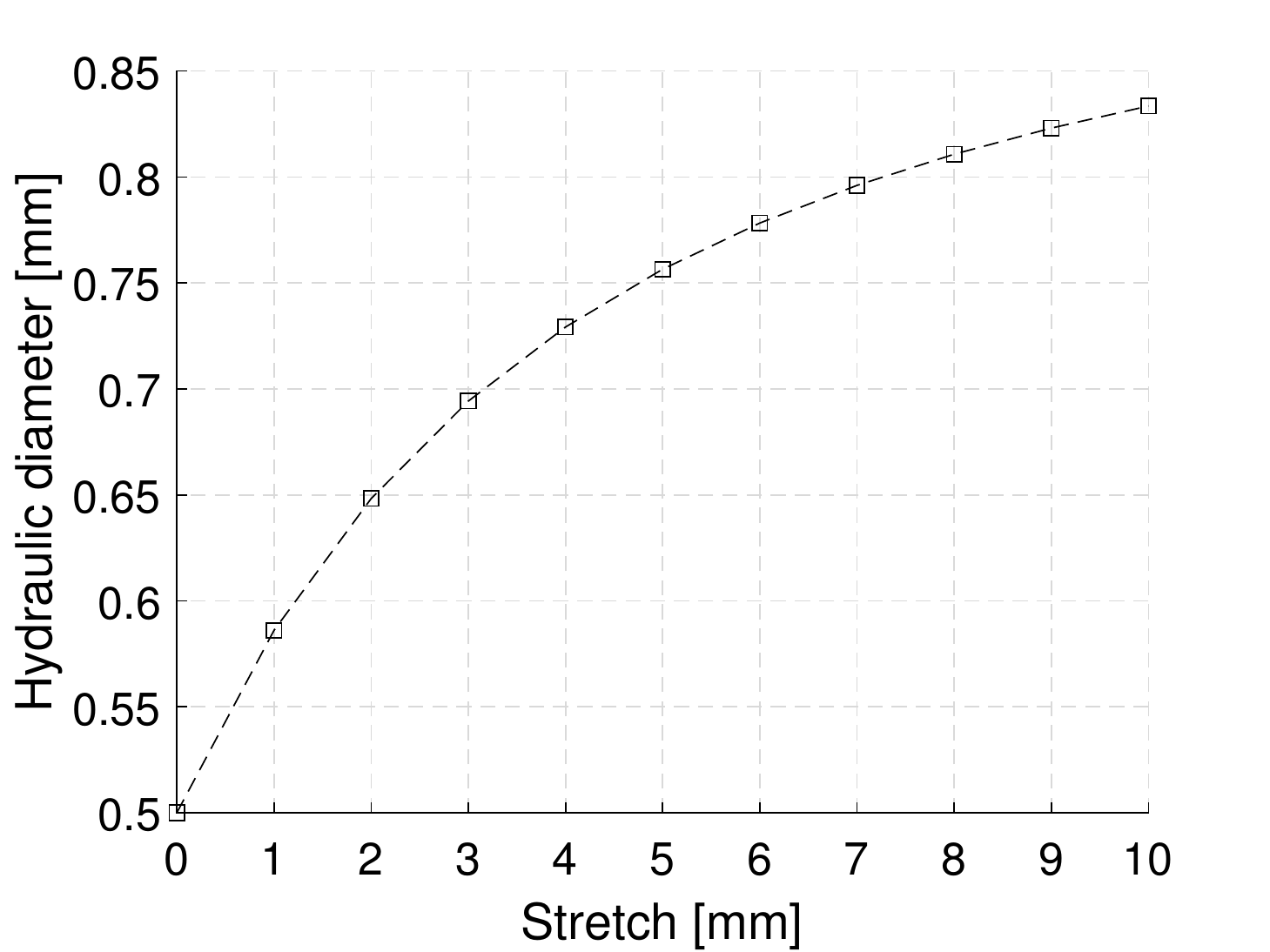}
				\caption{}
				\label{fig:fig7a}
			\end{subfigure}
			~\quad
			\begin{subfigure}[t]{0.30\textwidth}
				\centering
				\includegraphics[scale=0.35]{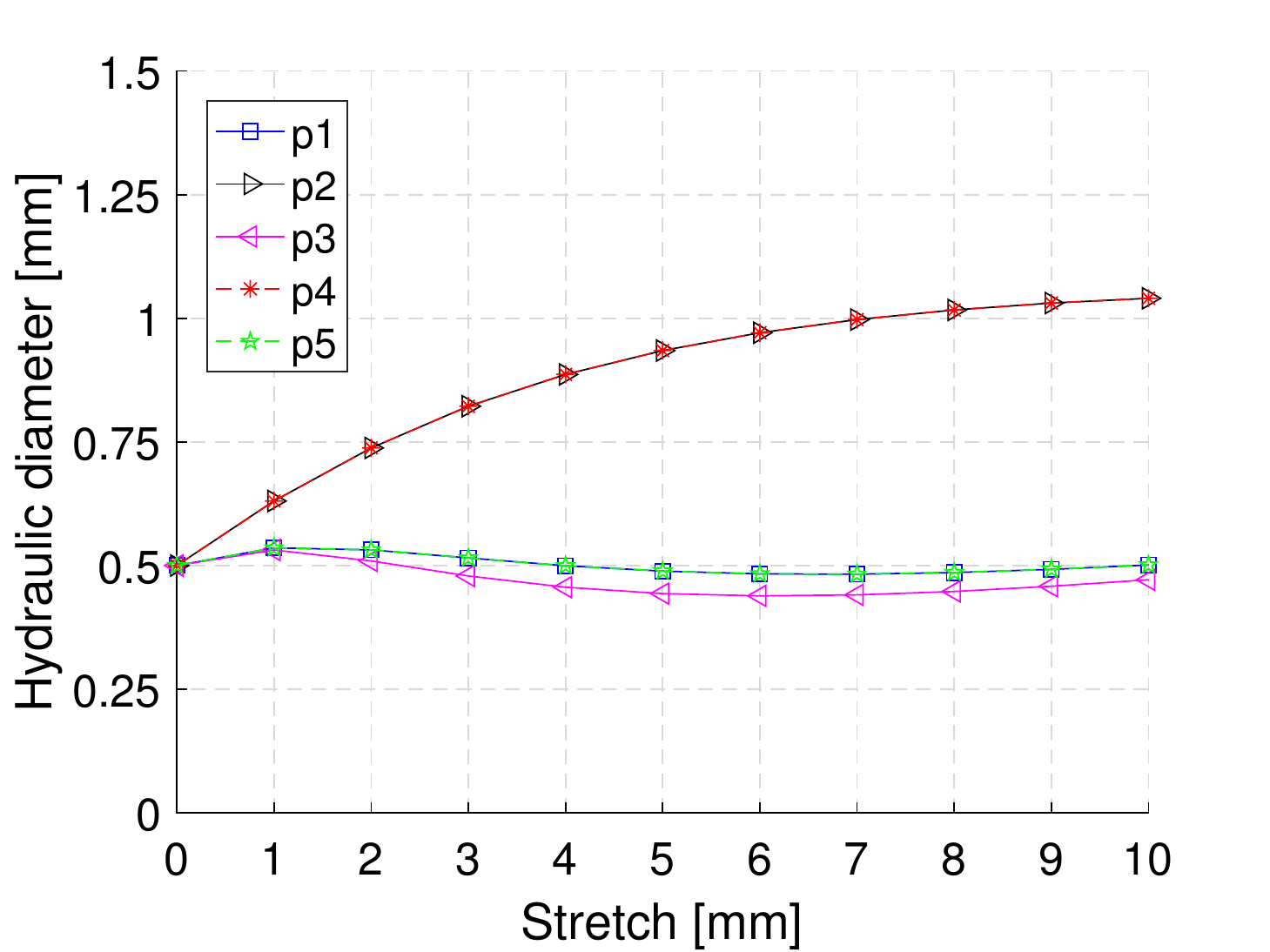}
				\caption{}
				\label{fig:fig7b}
			\end{subfigure}
			~\quad
			\begin{subfigure}[t]{0.30\textwidth}
				\centering
				\includegraphics[scale=0.35]{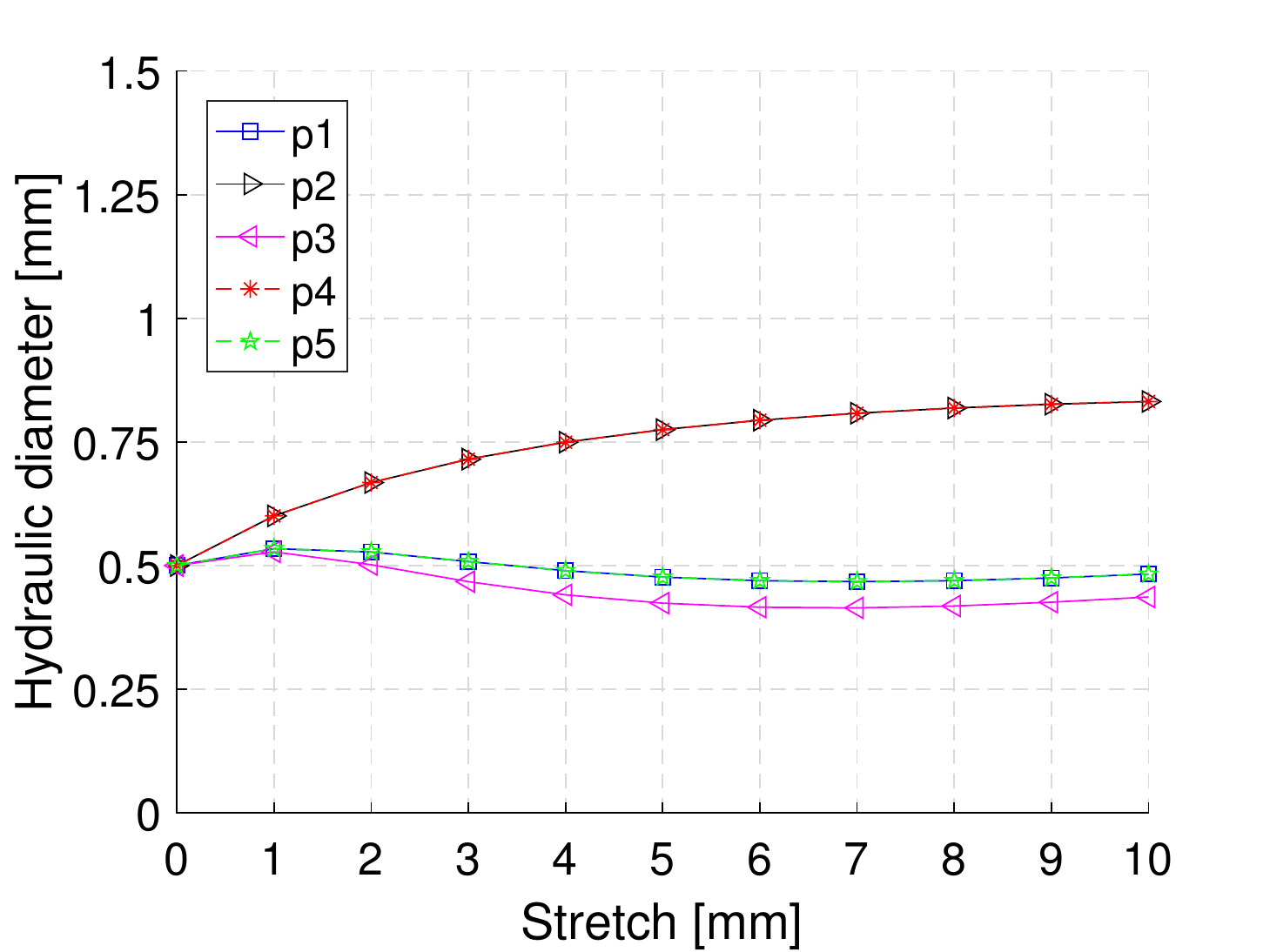}
				\caption{}
				\label{fig:fig7c}
			\end{subfigure}
			\caption{(\subref{fig:fig7a}) A plot for hydraulic diameter of a central pore for a $0.5$ mm thick flat membrane with respect to  stretching. Variation of hydraulic diameters with respect to the stretch for the different functional pores of CFCS IV obtained using weighted objective, (\subref{fig:fig7b}) with weights $w_1 = w_3 = w_5 = 5$ and $w_2 = w_4 = 1$ (\subref{fig:fig7c}) with weights $w_1 = w_3 = w_5 = 1$ and $w_2 = w_4 = 0$.}\label{fig:fig7}
		\end{figure}

		For the optimized CFCS IV structure, it is desired that pores p$_1$, p$_3$, and p$_5$ contract while pores p$_2$ and p$_4$ must be enlarged as the continuum deforms in response to an external stretching. However, it is noticed (Fig. \ref{fig:fig6l}) that the hydraulic diameters of the pores p$_2$ and p$_4$  increase, which is desired. Further, the hydraulic diameters of pores p$_1$ and p$_5$ gained slightly whereas a  minor reduction in the hydraulic diameter of pore p$_3$ can be observed. Therefore, a discrepancy can be noted in the behaviour of pores p$_1$ and p$_5$. In addition, the local topographical features around those pores are not same as those obtained for CFCS II. The cardinal reason could be the way objective $f_0$ is formulated wherein sum of squared differences between actual hydraulic diameters and their respective targets involves equal weights for enlarging and contracting output responses. In addition, a central pore of a flexible flat sheet made of PDMS gradually enlarges (Fig. \ref{fig:fig7a}) as the structure deforms, which implies that to contract a pore, proper topographical features are essential. Further, the optimizer has comparatively less design area available between two juxtaposed embedded pores of CFCS IV to relocate material in the required fashions to facilitate the desired contraction of the pores. Note also that, in comparison to this flat sheet, CFCS I reaches twice the final hydraulic diameter and the pore of CFCS II contracts an approximately four times lower value, which illustrates the  benefit of the optimized topography. An immediate treatment could be to increase the weights associated with the contracting pore(s) while evaluating the objective $f_0$. CFCS IV is resolved by extremizing the weighted (modified) objective $f_o^m$
		\begin{equation}\label{Eq:29}
		f_0^m = \displaystyle\sum_{i=1}^{nfp}w_{i}\bigg(D_i^* - D_i\bigg)^2
		\end{equation}
		with weights $w_{i}|_{i = 1,\,2,\,\cdots,\, 5}$ for pores p$_i|_{i = 1,\,2,\cdots,\,5}$; wherein $w_1 = w_3 = w_5 = 5$ and $w_2 = w_4 = 1$ are set.
		
		 For this case, a plot for hydraulic diameters of the different pores is depicted in Fig. \ref{fig:fig7b}. One can notice,   (gradual) hydraulic diameters of the pores p$_1$, p$_3$, and p$_5$ decrease as the structure is being actuated. However, no significant reduction is observed compared to the previous CFCS IV case (Fig. \ref{fig:fig6k}).  The hydraulic diameters of pores p$_2$ and p$_4$ are increasing, however their maximum values are less compared to that observed in the previously designed CFCS IV (Figs. \ref{fig:fig6k} and \ref{fig:fig7b}). Further, with weights $w_1 =w_3= w_5 = 1,\,w_2 = w_4 = 0$ (extreme values of the weights), the contraction and enlargement of the pores are depicted via variation in their corresponding hydraulic diameters with respect to stretching in Fig. \ref{fig:fig7c}. One notices that the hydraulic diameters of the pores p$_1$, p$_3$ and p$_5$ decrease from their initial states but not considerably, i.e., the pores contract but not significantly. Near maximum stretching, the hydraulic diameters of these pores are found to gradually increase again. On the other hand, pores p$_2$ and p$_4$ consistently enlarge under stretching. For this asymmetry, apart from the aforementioned reasons related to the natural tendency (Fig. \ref{fig:fig7a}) of pore(s) within a PDMS sheet under stretching and availability of less design area between pores of CFCS IV (Fig. \ref{fig:fig5c}) for the optimizer to relocate material to achieve the desired functionality, the following could also be a reason. In the current optimization process for the considered CFCSs, targets have only been imposed on the end equilibrium state at maximum stretching, instead of prescribing the entire deformation curve.  Therefore, it is certainly possible that an enlarging trend (positive slope, matching the natural tendency under stretching) is seen in the (almost) fully stretched states.

		\subsection{Prototypes of CFCSs and their performances}\label{section:4.2}

\begin{figure}[H]
		\centering
		\begin{subfigure}{0.45\textwidth}
			\includegraphics[scale=0.5]{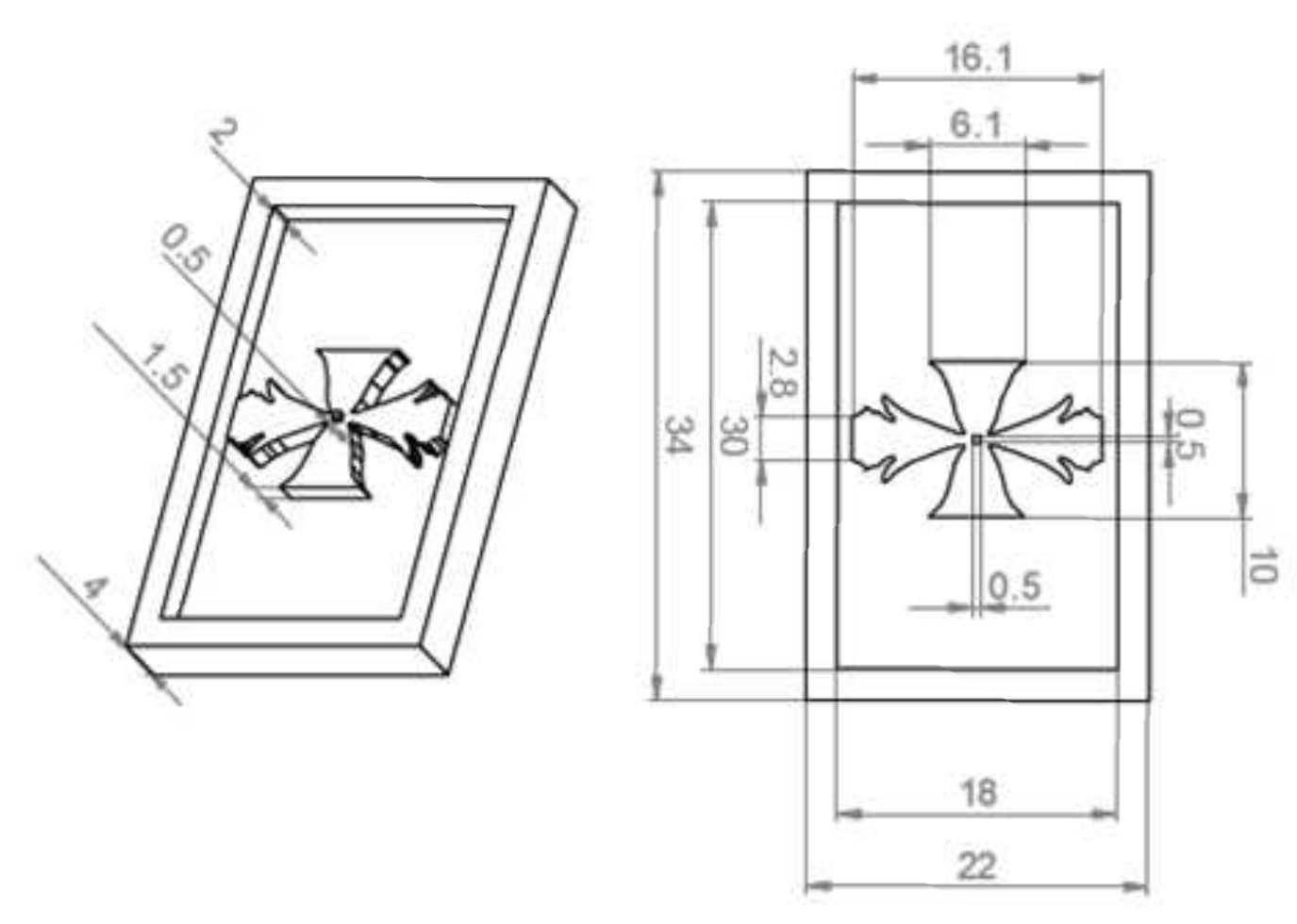}
			\caption{}
			\label{fig:fig8a}
		\end{subfigure}
		\begin{subfigure}{0.45\textwidth}
			\includegraphics[scale =0.5]{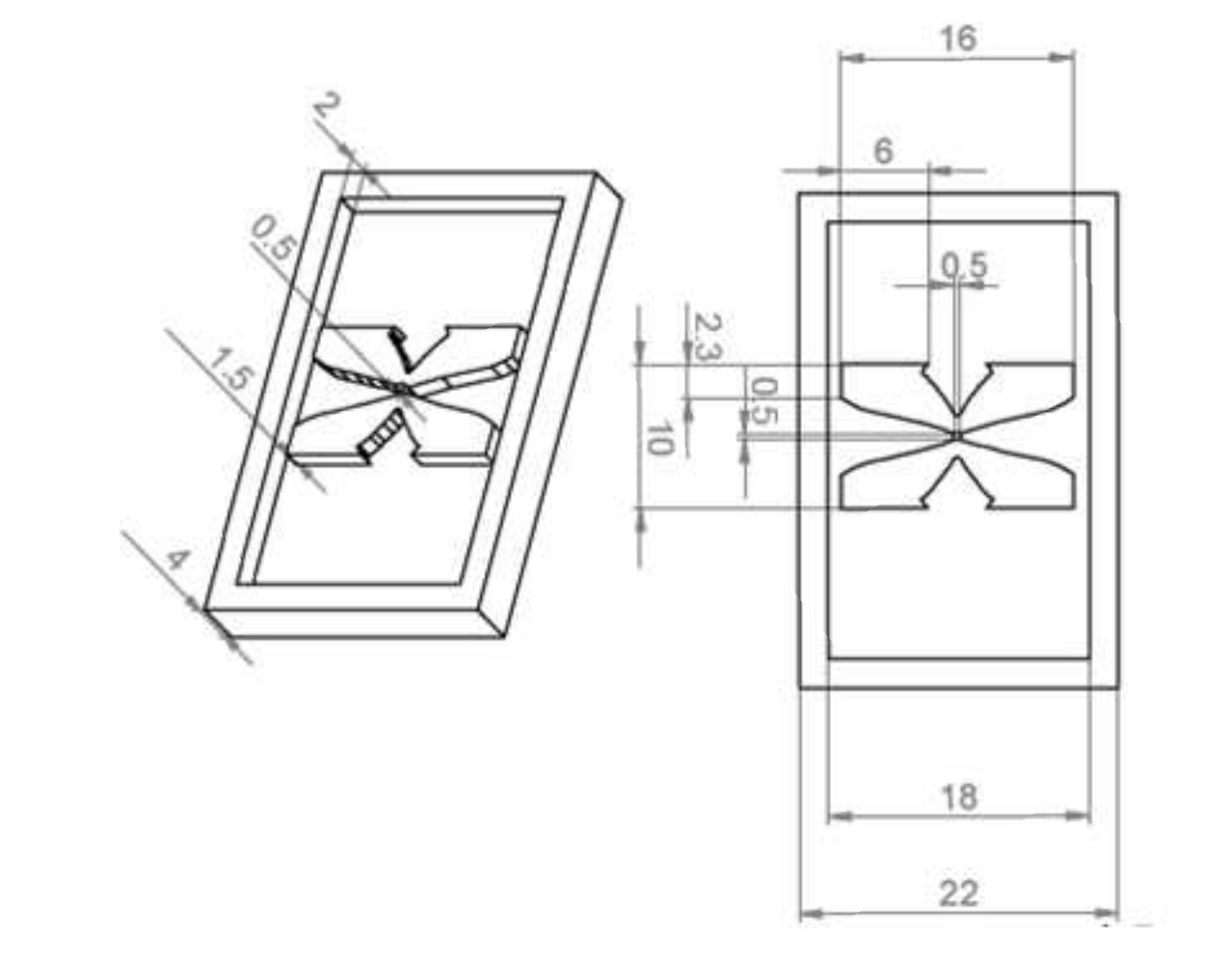}
			\caption{}
			\label{fig:fig8b}
		\end{subfigure}
		\caption{(\subref{fig:fig8a}) Mold for CFCS I, (\subref{fig:fig8b}) Mold for CFCS II. All dimensions are in mm.}\label{fig:fig8}
	\end{figure}

\begin{figure}[h!]
			\centering
			\begin{subfigure}{0.45\textwidth}
				\includegraphics[scale=1]{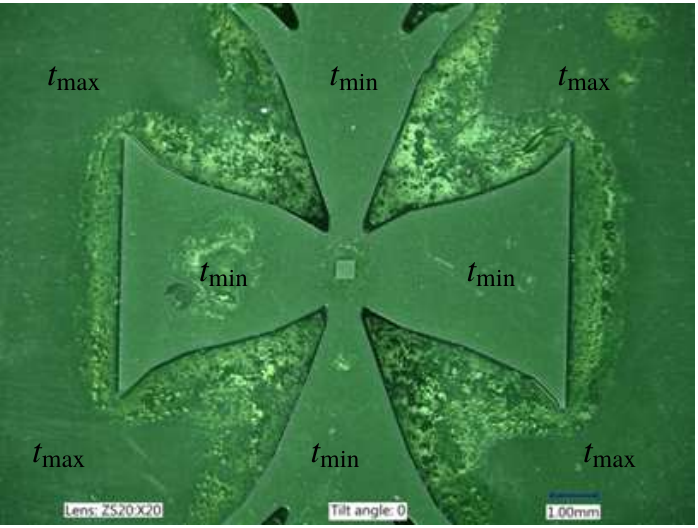}
				\caption{Prototype: CFCS I}
				\label{fig:fig9a}
			\end{subfigure}
		\qquad
			\begin{subfigure}{0.45\textwidth}
				\includegraphics[scale =1]{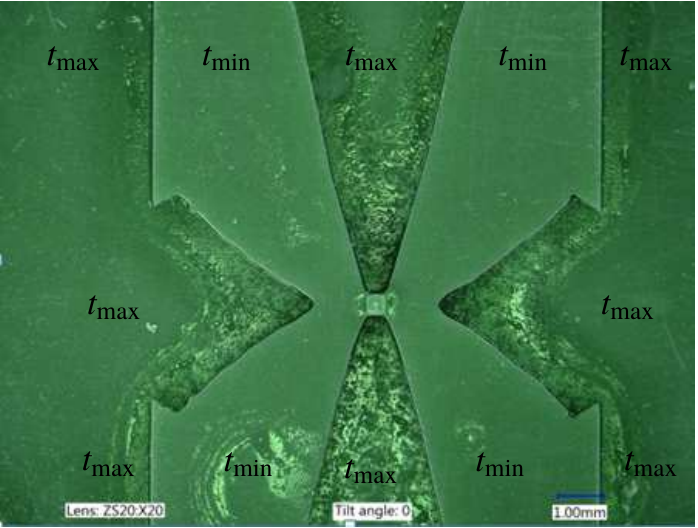}
				\caption{Prototype: CFCS II}
				\label{fig:fig9b}
			\end{subfigure}
			\caption{Prototypes for CFCS I and CFCS II, fabricated using PDMS}\label{fig:fig9}
		\end{figure}

\begin{figure}[h!]
  	\centering
  	\begin{subfigure}{0.450\textwidth}
  		\includegraphics[scale=0.7]{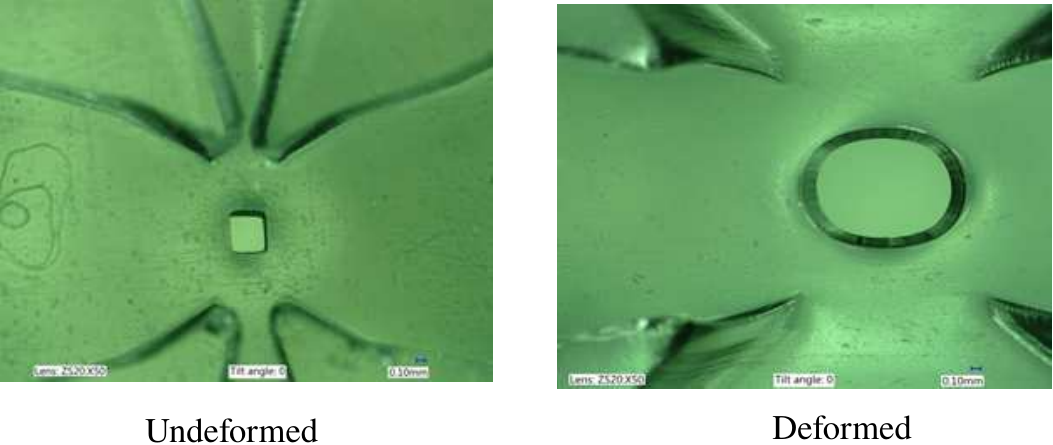}
  		\caption{}
  		\label{fig:fig10a}
  	\end{subfigure}
  	\qquad
  	\begin{subfigure}{0.450\textwidth}
  		\includegraphics[scale =0.7]{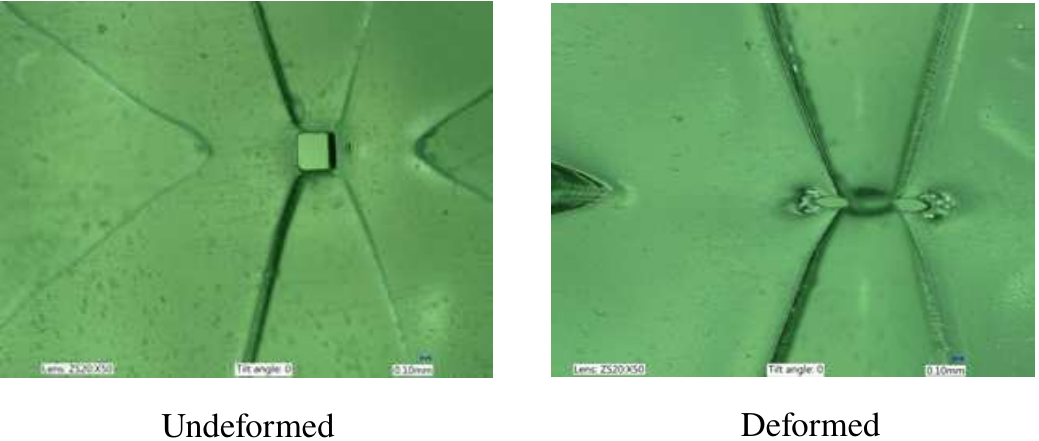}
  		\caption{}
  		\label{fig:fig10b}
  	\end{subfigure}
  	\caption{(\subref{fig:fig10a}) depicts the configurations of pore of CFCS I at undeformed and deformed states. (\subref{fig:fig10b}) shows the initial and final state of the configurations of CFCS II pore at undeformed and deformed states.}\label{fig:fig10}
  \end{figure}

	     CFCS I and CFCS II are fabricated using PDMS (1:10) and the technique presented in Section \ref{FT}. The dimensions of the mold designs for fabricating the structures are shown in Fig. \ref{fig:fig8}. The final prototypes for these continua are depicted in Fig. \ref{fig:fig9}. Some simplifications of fine structural features of the designs have been applied to facilitate release from the molds. The device depicted in Fig. \ref{fig:fig2c} is used to apply the stretching displacement. Images of the CFCS at different deformation states have been taken with a Keyence Digital Microscope VHX-6000. These images are then analyzed with the software ImageJ to obtain the area and perimeter of the pore and thus, the corresponding hydraulic diameter (Eq. \ref{Eq:23}) is evaluated. A close view of the different states of functional pores of the CFCS I and CFCS II in undeformed and deformed configurations is  depicted via Fig. \ref{fig:fig10}. In case where the pore is contracting, it can be noticed (Fig. \ref{fig:fig10b}) that boundary of the pores eventually comes into contact and thus, demanding a treatment for self contact \cite{Kumar2018} behaviour in the synthesis approach. This is an additional challenge which is left for future work.  
		
	\begin{figure}[h!]
		\centering
		\begin{subfigure}{0.45\textwidth}
			\includegraphics[scale=0.5]{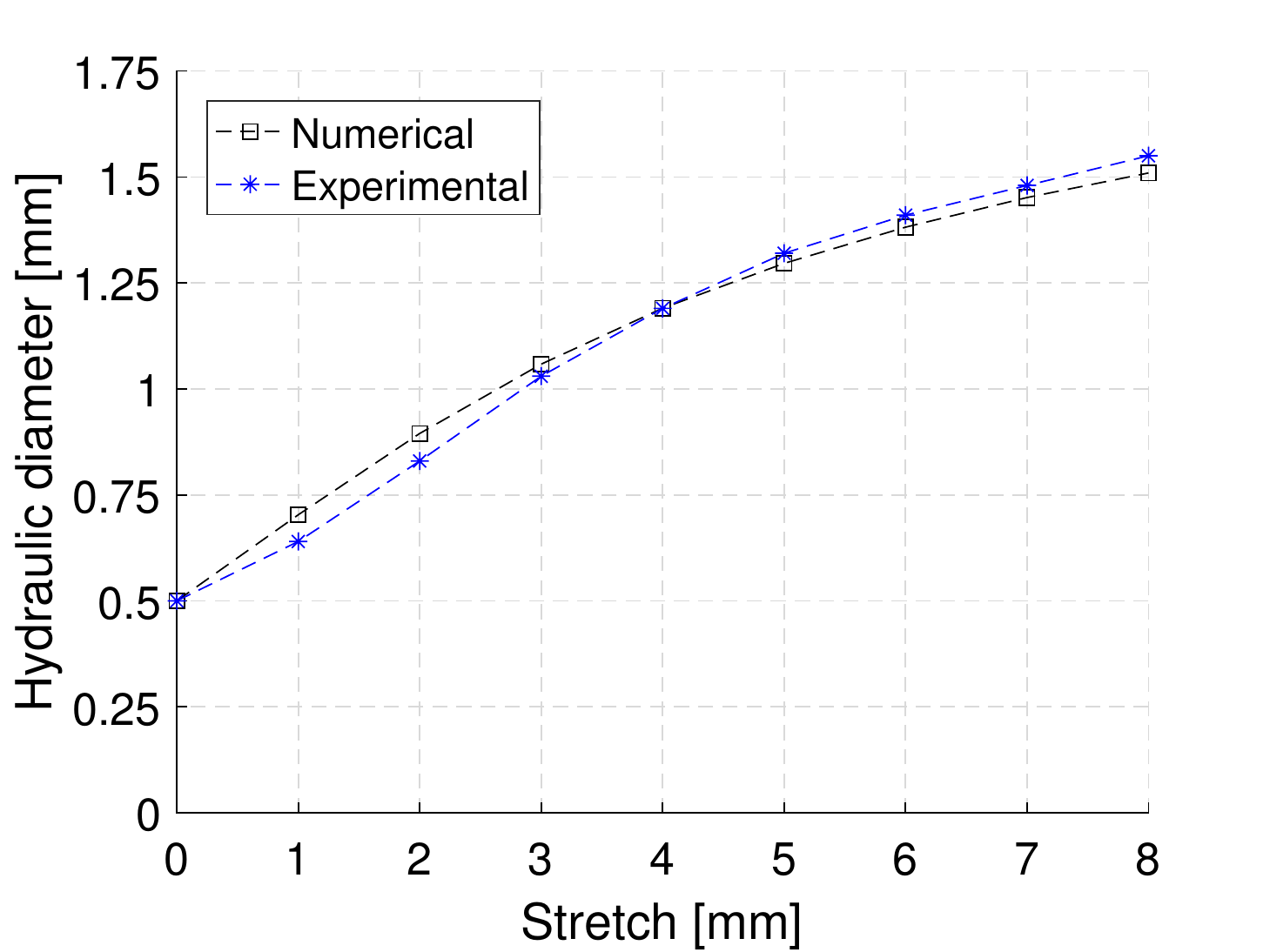}
			\caption{}
			\label{fig:fig11a}
		\end{subfigure}
		\qquad
		\begin{subfigure}{0.45\textwidth}
			\includegraphics[scale =0.5]{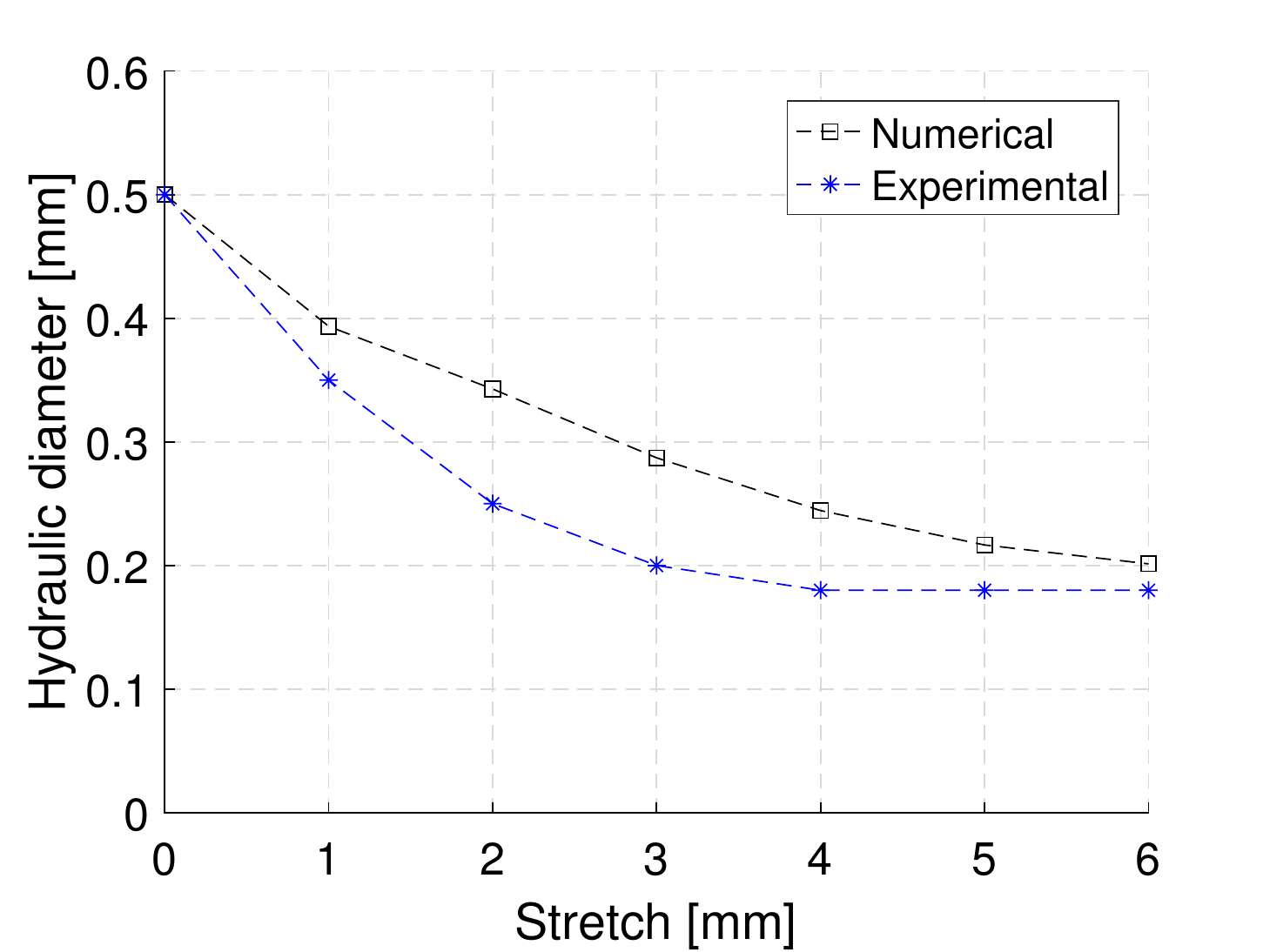}
			\caption{}
			\label{fig:fig11b}
		\end{subfigure}
		\caption{Experimental and numerical hydraulic diameters for the pores of CFCS I (\subref{fig:fig11a}) and CFCS II (\subref{fig:fig11b}) at different instances of stretching }\label{fig:fig11}
	\end{figure}

	One notices (Fig. \ref{fig:fig11a}) that hydraulic diameters of the pore obtained from experimental and numerical results are in good agreement with each other for CFCS I. For the contracting pore of CFCS II, a minor deviation can be observed initially, however the trend is almost similar. Given the inaccuracies that may have occurred in manufacturing and testing, the correlation between experimental and numerical results is encouraging.
\section{Conclusions}\label{section:5}
	The concept and synthesis approach for planar Compliant Fluidic Control Structures which are monolithic flexible continua with a number of functional pores, is presented. It can be seen that the presented approach can successfully generate such structures for various desired responses for their functional pores. In millimeter range, these structures are planar, in general.  Therefore, while using geometrical and material nonlinearity within the synthesis approach plane-stress conditions are imposed. The considered planar CFCSs consist of different topographical features and to determine those, an optimization problem is formulated in terms of thickness design variables. The structures are subjected to large deformations, with strains up to 100\%. Nevertheless no numerical instabilities have been observed.
	
     To evaluate the regulated fluid flow control, hydraulic diameters of the pores are used. An objective is formed using target values for the hydraulic diameter of each pore, corresponding to the desired enlarging or contracting behavior. Structures controlling multiple pores simultaneously can be generated. The optimized CFCS designs have been fabricated using PDMS material.  The experimental and numerical results are in good agreement indicating that the Arruda-Boyce material definition can effectively represent the constitutive behavior of PDMS for this situation. In future, the work can be directed towards 3D TO, and other manufacturing techniques, e.g.,  additive manufacturing, can be employed to fabricate the optimized 3D CFCSs. In case of contracting a pore, the boundary defining it eventually comes into contact. The effect of such contact constraints also forms a future challenge.

		\begin{appendices}
			\section{The Cauchy Stress Tensor and Elastic Tensor}\label{Append:A}
		In view of Eqs. \ref{Eq:11}-\ref{Eq:13}, we have
		\begin{equation}\label{Eq:30}
		\Psi = a_1(\bar{\mathcal{J}}_{1\bm{B}} - 3) + a_2(\bar{\mathcal{J}}_{1\bm{B}}^2 -9) + a_3(\bar{\mathcal{J}}_{1\bm{B}}^3 -27) + \frac{K}{2}\bigg[\frac{J^2-1}{2} - \ln J\bigg].
		\end{equation}
		From the fundamentals of continuum mechanics \cite{holzapfel2001nonlinear}, we know that
		\begin{equation}\label{Eq:31}
		\frac{\partial \mathcal{J}_{1\bm{B}}}{\partial \bm{B}} = \bm{I}, \, \, \frac{\partial J}{\partial \bm{B}} = \frac{1}{2}J\bm{B}^{-1},\,\,\bm{\sigma} = \frac{2}{J}\frac{\partial \Psi}{\partial \bm{B}} \bm{B},\,\,\mathbb{C} = \frac{4}{J}\bm{B}\frac{\partial^2\Psi}{\partial\bm{B}^2}\bm{B}.
		\end{equation}
		Since, $\bar{\mathcal{J}}_{1\bm{B}} = J^{-\frac{2}{3}}\mathcal{J}_{1\bm{B}}$. Therefore, using Eq. \ref{Eq:31} and applying the chain rules of differentiation, one can have
		\begin{equation}\label{Eq:32}
			\frac{\partial \bar{\mathcal{J}}_{1\bm{B}}}{\partial \bm{B}} = J^{-\frac{2}{3}}\bigg[\bm{I} - \frac{1}{3}\bm{B}^{-1}\mathcal{J}_{1\bm{B}}\bigg]
		\end{equation}
		and
		\begin{equation}\label{Eq:33}
		\frac{\partial^2 \bar{\mathcal{J}}_{1\bm{B}}}{\partial \bm{B}^2} = J^{-\frac{2}{3}}\bigg[\frac{1}{9}\bm{B}^{-1}\otimes\bm{B}^{-1}\mathcal{J}_{1\bm{B}} + \frac{1}{3}\bm{B}^{-1}\cdot\mathbb{I}\cdot\bm{B}^{-1} \mathcal{J}_{1\bm{B}} - \frac{1}{3}\bm{B}^{-1}\otimes\bm{I}-\frac{1}{3}\bm{I}\otimes\bm{B}^{-1} \bigg]
		\end{equation}
		where $\mathbb{I}$ is a fourth order symmetric tensor. 
		Now, in view of Eqs. (\ref{Eq:30}-\ref{Eq:33}), using the chain rules of differentiation, one can find  the Cauchy stress tensor as
		\begin{equation}\label{Eq:34}
		\bm{\sigma} = 2J^{-\frac{5}{3}}(a_1 + 2a_2\bar{\mathcal{J}}_{1\bm{B}} + 3a_3{\bar{\mathcal{J}}_{1\bm{B}}}^2)\bigg[\bm{B} - \frac{1}{3}\mathcal{J}_{1\bm{B}}\bm{I}\bigg] + \frac{K}{2J}(J^2 -1)\bm{I}
		\end{equation}
		and the material tangent tensor as
		\begin{equation}\label{Eq:35}
		\mathbb{C} = \frac{4}{J}(a_1 + 2a_2\bar{\mathcal{J}}_{1\bm{B}} + 3a_3{\bar{\mathcal{J}}_{1\bm{B}}}^2)\bm{T}_1 +  \frac{4}{J}(2a_2 + 6a_3{\bar{\mathcal{J}}_{1\bm{B}}})\bm{T}_2 + K\bigg[\bm{I}\otimes\bm{I} -\mathbb{I}\bigg]J + \frac{K}{J}\mathbb{I}
		\end{equation}
		where
		\begin{equation}\label{Eq:36}
		\bm{T}_1 = J^{-\frac{2}{3}}\bigg[\frac{1}{9}\mathcal{J}_{1\bm{B}}\bm{I}\otimes\bm{I} + \frac{1}{3}\mathcal{J}_{1\bm{B}}\mathbb{I} - \frac{1}{3}\bm{B}\otimes\bm{I}-\frac{1}{3}\bm{I}\otimes\bm{B} \bigg]
		\end{equation}
		and
		\begin{equation}\label{Eq:37}
		\bm{T}_2 = J^{-\frac{4}{3}}\bigg[\bm{B}\otimes\bm{B} - \frac{1}{3}\mathcal{J}_{1\bm{B}}\bm{B}\otimes\bm{I}-\frac{1}{3}\mathcal{J}_{1\bm{B}}\bm{I}\otimes\bm{B}+\frac{1}{9}\mathcal{J}_{1\bm{B}}^2\bm{I}\otimes\bm{I}\bigg].
		\end{equation}
	\section{Convergence history plot}\label{Append:B}	
	\begin{figure}[h!]
			\centering
			\includegraphics[scale = 0.85]{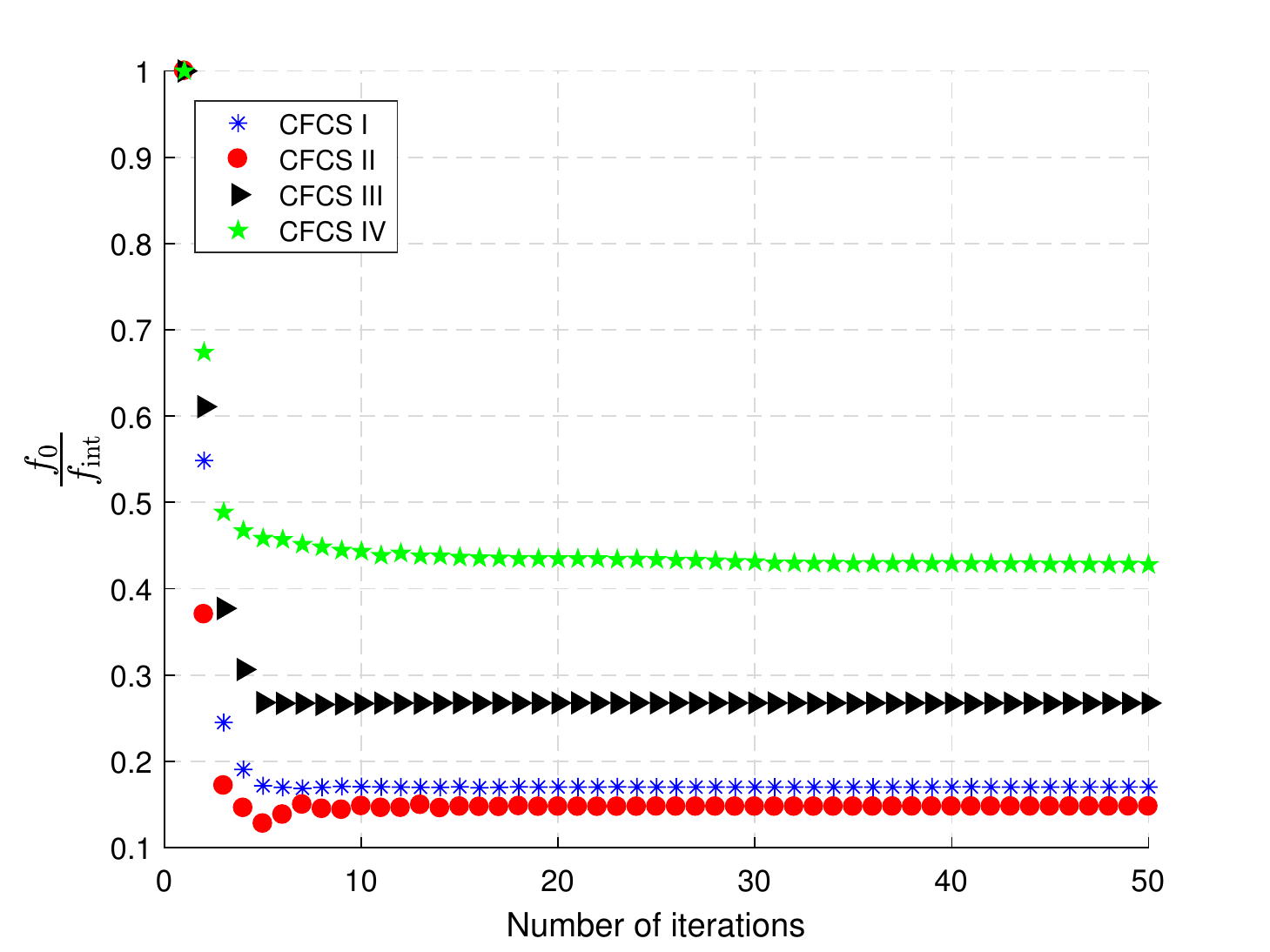}
			\caption{Convergence histories of CFCSs. $f_{\mathrm{int}}$ is value of the objective at  the first optimization iteration.}
			\label{fig:fig12}
    \end{figure}
		\section{Internal Force and Stiffness Matrix}\label{Append}
			The elemental internal force vector acting on an element $\Omega_e$ can be calculated as
			\begin{equation}\label{Eq:47}
			\mathbf{f}^e_\mathrm{int} = \int_{\Omega_e}\mathbf{B}_{UL}^T\bm{\sigma} \mathrm{d}\text{v}; \qquad \mathbf{B}_{UL}^a = \bigg[\mathbf{B}^1_{UL}\,\,\mathbf{B}^2_{UL}\,\, \cdots\,\, \mathbf{B}^{n}_{UL} \bigg]
			\end{equation}
			where
		     \begin{equation}\label{Eq:48}
		      \bm{B}^a_\mathrm{UL}=
		      \begin{bmatrix}
		        N_{a,1} & 0 \\
		       0 & N_{a,2} \\
		     N_{a,2} &  N_{a,1}
		     \end{bmatrix}, \qquad
		    \frac{\partial N}{\partial \bm{x}}=
		      \begin{bmatrix}
		      N_{a,1} \\
		      N_{a,2} 
		\end{bmatrix}
		\end{equation}
		\end{appendices}
		and $n$ is the number of nodes in an element $\Omega_e$.
		
		One evaluates elemental stiffness matrix $\mathbf{K}_\mathrm{int}^e$ as
		\begin{equation}\label{Eq:49}
		\mathbf{K}_\mathrm{int}^e = \mathbf{K}_\mathrm{mat}^e + \mathbf{K}_\mathrm{geo}^e
		\end{equation}
		where $\mathbf{K}_\mathrm{mat}^e\,\text{and}\,\mathbf{K}_\mathrm{geo}^e$ are contribution from  material and geometrical nonlinearities, respectively. These can be evaluated as
		\begin{equation}\label{Eq:50}
		\mathbf{K}_\mathrm{mat}^e = \int_{\Omega^e}\mathbf{B}_{UL}^T \mathbb{C}\mathbf{B}_{UL}dv
		\end{equation}
		 where $\mathbb{C}$ is the material tangent/elastic modulus (see Appendix \ref{Append:A}) and
		\begin{equation}\label{Eq:51}
		\mathbf{K}_\mathrm{geo}^e = 
		\begin{bmatrix}
		\mathbf{K}_\mathrm{geo}^{11} & \mathbf{K}_\mathrm{geo}^{12} & \cdots & \mathbf{K}_\mathrm{geo}^{1n}\\
		\mathbf{K}_\mathrm{geo}^{21} & \mathbf{K}_\mathrm{geo}^{22} & \cdots & \mathbf{K}_\mathrm{geo}^{2n}\\
		\vdots & \vdots & \ddots &\vdots \\
		\mathbf{K}_\mathrm{geo}^{n1} & \mathbf{K}_\mathrm{geo}^{n2} & \cdots & \mathbf{K}_\mathrm{geo}^{nn}
		\end{bmatrix}, \quad \mathbf{K}_\mathrm{geo}^{ab}  = \mathbf{I}\int_{\Omega^e} {N}_{a,i} \sigma_{ij} N_{b,j}dv
		\end{equation} 
		 In the expression of $\mathbf{K}_\mathrm{geo}^{ab}$, $i$ and $j$ are dummy summation indices and $\mathbf{I}$ is the identity matrix in $\mathcal{R}^d$.
		
		To perform numerical integration, one transforms the integrals from the elemental $\Omega_e$ to Gaussian range $\bm{\xi}\in[-1\, \, 1]$, e.g.,
		\begin{equation}\label{Eq:52}
		\begin{aligned}
		\int_{\Omega^e} f(\bm{x})dv = \int_\square {f}(\bm{\xi})\det j(\bm{\xi})d\square &= t_e(\zeta_e)\int_{\xi_1,\,\xi_2}{f}(\bm{\xi})\det j(\bm{\xi})d\xi_1 d\xi_2\\ &=t_e(\zeta_e)\sum_{gp=1}^{n_{gp}}{f}(\bm{\xi}_{gp})\det j(\bm{\xi}_{gp})w_{gp}
		\end{aligned}
		\end{equation} 
		where $\det j(\bm{\xi})$ is determinant of the Jacobian for the coordinate system considered, $w_{qp}$ is the weight factor at integration point $qp$, $\square$ indicates the range of parent coordinates for the dimension of problem considered and $t_e(\zeta_e)$ is the thickness (Eq. \ref{Eq:1}) of the planar membrane. Now, in view of Eq. (\ref{Eq:52}), one evaluates $\mathbf{f}_\mathrm{int}^e$, $\mathbf{K}_\mathrm{mat}^e$ and $\mathbf{K}_\mathrm{geo}^{ab}$ as
		\begin{equation}\label{Eq:53}
			\mathbf{f}^e_\mathrm{int} = t_e(\zeta_e)\sum_{gp=1}^{n_{gp}}\mathbf{B}_{UL}^T(\bm{\xi}_{gp})\bm{\sigma}(\bm{\xi}_{gp})\det j(\bm{\xi}_{gp}) w_{gp}
		\end{equation}
		\begin{equation}\label{Eq:54}
		\mathbf{K}_\mathrm{mat}^e = t_e(\zeta_e)\sum_{gp=1}^{n_{gp}}\mathbf{B}_{UL}^T(\bm{\xi}_{gp}) \mathbb{C}(\bm{\xi}_{gp})\mathbf{B}_{UL}(\bm{\xi}_{gp})\det j(\bm{\xi}_{gp}) w_{gp}
		\end{equation}
		\begin{equation}\label{Eq:55}
		\mathbf{K}_\mathrm{geo}^{ab} = t_e(\zeta_e)\sum_{gp=1}^{n_{gp}}{N}_{a,i}(\bm{\xi}_{gp}) \sigma_{ij}(\bm{\xi}_{gp}) N_{b,j}(\bm{\xi}_{gp})\det j(\bm{\xi}_{gp}) w_{gp}
		\end{equation}
		\bibliography{myreference}
		\bibliographystyle{ieeetr}
		 \end{document}